\documentclass[conference]{IEEEtran}

\usepackage{tikz}
\usepackage{pgfplots}
\pgfplotsset{compat=newest} 
\pgfplotsset{plot coordinates/math parser=false}
\usepackage{graphicx}
\usepackage{color}
\usepackage[numbers, square, comma, sort&compress]{natbib}
\usepackage{amsmath,amsthm}
\usepackage{amssymb, bm, dsfont}
\usepackage{float}
\usepackage{amssymb,stmaryrd,amsmath,amsfonts,rotating}
\usepackage{booktabs}

\usepackage[ruled,vlined]{algorithm2e}

\newtheorem{theorem}{Theorem}

\newtheorem{lemma}{Lemma}
\newtheorem{corollary}[theorem]{Corollary}

\theoremstyle{definition}
\newtheorem{remark}{Remark}
\newtheorem{definition}{Definition}
\ifCLASSINFOpdf
\else
\fi



\makeatletter
\newcommand*\titleheader[1]{\gdef\@titleheader{#1}}
\AtBeginDocument{%
  \let\st@red@title\@title%
  \def\@title{%
    \bgroup\normalfont\large\centering\@titleheader\par\egroup
    \vskip1.5em\st@red@title}
}
\makeatother

\title{\vspace*{-6mm}Nonlinear Fourier Transform of Truncated Multi-Soliton Pulses}

\author{\IEEEauthorblockN{Vahid Aref}
\IEEEauthorblockA{Nokia Bell Labs, Stuttgart, Germany 
\\Email: vahid.aref@nokia-bell-labs.com
}
}

\titleheader{\vspace*{-15mm}\footnotesize This paper has been accepted for presentation at 
12th ITG Conference on Systems, Communications and Coding (SCC) 2019, Feb. 2019. The link to the final version in the Proceedings of SCC will be posted here once it is available. }

\begin{document}
\maketitle


\begin{abstract}
Multi-soliton pulses, as special solutions of the Nonlinear Schr{\"o}dinger 
Equation (NLSE), are potential candidates for  optical fiber transmission
where the information is
modulated and recovered in the so-called nonlinear Fourier domain.
For data communication, the exponentially decaying tails of a multi-soliton must be truncated.
Such a windowing changes the nonlinear Fourier spectrum of the pulse.
The results of this paper are twofold: $(i)$ we derive the simple closed-form expressions for the nonlinear
spectrum, discrete and continuous spectrum, of a symmetrically truncated multi-soliton pulse from tight approximation of the truncated tails. 
We numerically show the accuracy of the closed-form expressions. $(ii)$ We show how to find, in general, 
the eigenvalues of the discrete spectrum from the continuous spectrum. 
We present this method for the application in hand. 
\end{abstract}


%
\IEEEpeerreviewmaketitle

\section{Introduction}

Multi-Soliton pulses are special solutions of the Nonlinear Schr{\"o}dinger Equation (NLSE), the basic model 
for the propagation of optical fields in the nonlinear optical fibers. Having a predictable evolution along the fiber makes solitons an attractive candidate for data modulation.
After the advent of coherent optical technology, the transmission capacities have been significantly increased to a point
that the Kerr nonlinearity becomes the limiting factor. The need 
for exploiting Kerr nonlinearity in system design attracts  the 
attention again to soliton pulses~\cite{yousefi2014information}. 
Moreover, the coherent technology allows utilizing all degrees of freedom of a soliton for modulation.

A multi-soliton pulse has a simple representation in a so-called nonlinear Fourier spectrum~\cite{shabat1972exact},\cite{ablowitz1981solitons}.
The spectrum has two distinct parts: Continuous spectrum and discrete spectrum. 
The continuous spectrum includes the real valued frequencies and is the counterpart of (linear) Fourier spectrum.
The discrete spectrum contains a set of isolated complex valued frequencies, called eigenvalues, representing the solitonic components of a pulse. A multi-soliton pulse is characterized with its $N$ eigenvalues with no continuous spectrum. 

As a signal propagates according to the NLSE, 
its complex spatial transformation in time domain can be characterized by simple transformations of its
nonlinear spectrum: the eigenvalues
(the complex eigenvalues and the real-valued frequencies)
are preserved and the spectral amplitude of each eigenvalue transforms 
\textit{independently} via a \textit{simple} transfer function. 
These properties motivate to modulate and/or detect data over nonlinear spectrum.
Such modulation schemes are named Nonlinear Frequency Division Multiplexing (NFDM)~\cite{yousefi2014information}.
The transmission of multi-solitons has been 
demonstrated 
with different modulation formats in various scenarios \cite{Buelow20167eigenvalues,aref2015experimental,hari2016multi,
gui2017alternative,aref2016onoff,dong2015nonlinear,gui2017high,
geisler2016nfdm,buchberger2018probabilistic,
shevchenko2018capacity,gaiarin2018dual}. 
Similar studies have been done for
modulation of continuous spectrum in
\cite{le2015nonlinear,le2017125gbps,civelli2017noise,
yousefi2016linear,Gemechu2018dual,Xianhe18dual} as well as 
modulation of both discrete and continuous spectrum in
\cite{le2017nature,aref2016condis,Tavakkolnia2015bothspectra}.
  
Transmitting a train of multi-soliton pulses, the discrete spectrum of each pulse is modulated independently. 
Since a multi-soliton pulse has exponentially decaying tails, the tails must be truncated properly to avoid inter-symbol interference and also to increase the symbol rate. However, the truncation perturbs the modulated discrete spectrum by changing the eigenvalues and their spectral amplitudes and it also generates a non-zero 
continuous spectrum. 
An aggressive truncation may also cause nonlinear 
interaction between adjacent pulses during the transmission. 
The effect of truncation is recently studied in  \cite{vaibhav2018exact} but no explicit expression is given and the distorted nonlinear spectrum must be computed numerically.

In this paper, we derive simple closed-form expressions for the nonlinear Fourier spectrum when the tails of a multi-soliton 
are symmetrically truncated. Although the analysis contains some tight approximation of the tails, the closed-form expression, for both continuous and discrete spectrum, seems to be numerically precise (for a non-aggressive truncation).
Our analysis is generic but we present it here for symmetric multi-solitons as 
the analysis and the final closed-form expression become simpler and these soltions are of practical interest.
We further show in general how to find the eigenvalues of the discrete spectrum from the continuous spectrum. We present this method for the application of this paper. The rest of the paper is organized as follows: the nonlinear Fourier spectrum and multi-soliton pulses are introduced in Section~\ref{sec:nft}.
The closed-form expressions are presented in Section~\ref{sec:trun} but their derivations are postponed to Section~\ref{sec:anal}. 
The analytic nonlinear spectrum is verified numerically in Section~\ref{sec:simul} and the paper is concluded in Section~\ref{sec:conclusion}.

\section{Nonlinear Fourier Spectrum and Multi-Solitons}\label{sec:nft}

The standard Nonlinear Schr{\"o}dinger Equation (NLSE) serves as
the basic model for the pulse propagation  $q(t,z)$ along an ideally lossless and noiseless fiber, 
\begin{equation}
\frac{\partial}{\partial z}q(t,z)+j\frac{\partial^2}{\partial t^2}q(t,z)+2j|q(t,z)|^2q(t,z)=0.
\label{NLSE}
\end{equation}
The physical pulse $Q(\tau,\ell)$ at location $\ell$ along the fiber is then described by
\begin{equation*}
Q\left(\tau,\ell\right)=\sqrt{P_0}\,\,\, q\left(\frac{\tau}{T_0},\ell \frac{\left|\beta_2\right|}{2T_0^2}\right) \text{ with } P_0\cdot T_0^2=\frac{\left|\beta_2\right|}{\gamma}, 
\end{equation*}
where $\beta_2<0$ is the chromatic dispersion and $\gamma$ is the Kerr nonlinearity of the fiber, and $T_0$ determines the symbol rate.  In practice, one should include the loss term, noise of amplifiers, higher order dispersion terms, \textit{etc.} in this simple model. 

\subsection{Nonlinear Fourier Transform}

The pulse propagation according to the NLSE can be characterized
by simple transformations in the nonlinear Fourier spectrum. Nonlinear Fourier Transform (NFT) maps a time-domain signal
$q\left(t,z\right)$ to its unique nonlinear spectrum by the following so-called Zakharov-Shabat system\footnote{The Zakharov-Shabat scattering problem is usually defined differently, e.g. \cite{ablowitz1981solitons}. This equivalent but simpler form is obtained by change of variables}~\cite{ablowitz1981solitons}

\begin{equation}\label{eq:ZS}
\frac{\partial}{\partial t}\!\left(\begin{matrix}v_1(t;z)\\ v_2(t;z)\end{matrix}\right)\!=\!
	\left(\begin{matrix}
	0 & \!\!\!\!\!\!\!\!q\left(t,z\right)e^{+2j\lambda t} \\-q^*\left(t,z\right)e^{-2j\lambda t} & \!\!0
	\end{matrix}\right)
	\left(\begin{matrix}v_1(t;z)\\ v_2(t;z)\end{matrix}\right),
\end{equation}
with the boundary condition
\begin{equation}\label{eq:boundary}
\lim_{t\to-\infty} \left(\begin{matrix} v_1(t;z)\\v_2(t;z)\end{matrix}\right)=\left(\begin{matrix}1\\0\end{matrix}\right).
\end{equation}
The nonlinear Fourier coefficients (Jost pair) are defined as
\begin{equation*}
\left(\begin{matrix}a\left(\lambda;z\right) \\ b\left(\lambda;z\right) \end{matrix}\right)=\lim_{t\to+\infty} \left(\begin{matrix} v_1(t;z)\\v_2(t;z)\end{matrix}\right).
\end{equation*}
An important property of the nonlinear spectrum
 is its simple linear evolution given by~\cite{ablowitz1981solitons}
\begin{align}
a\left(\lambda;z\right)&=a\left(\lambda\right),\nonumber\\
b\left(\lambda;z\right)&=b\left(\lambda\right)\exp(-4j\lambda^2z),\label{eq:b_evol}
\end{align} 
where we define $a\left(\lambda\right)=a\left(\lambda;z=0\right)$ and $b\left(\lambda\right)=b\left(\lambda;z=0\right)$.

The set $\Omega$ of isolated complex values, 
is the set of simple roots 
of $a(\lambda;z)$ with positive imaginary part, 
which are called \textit{eigenvalues} as they do not change 
in terms of $z$, i.e. $\lambda_k(z)=\lambda_k$. 
We denote the imaginary part of $\lambda$ by $\text{Im}(\lambda)$.
The nonlinear spectrum is usually described by the following two parts:

\begin{itemize}
\item[(i)] Continuous Part: the spectral amplitude $Q_c(\lambda;z)=b(\lambda;z)/a(\lambda;z)$ for real frequencies $\lambda\in\mathbb{R}$.
\item[(ii)] Discrete Part: $\{\lambda_k,Q_d(\lambda_k;z)\}$ where 
$\lambda_k\in\Omega$, i.e. $a(\lambda_k;z)=0$, and
$Q_d(\lambda_k;z)=b(\lambda_k;z)/\frac{\partial a(\lambda;z)}{\partial \lambda}|_{\lambda=\lambda_k}$
\end{itemize}

Note that there are several methods to compute the nonlinear spectrum by numerically solving  
the Zakharov-Shabat system. Some of these methods are summarized in \cite{turitsyn2017nonlinear,yousefi2014information,wahls2015fast}.

\begin{remark}
Because of the relation of \eqref{eq:b_evol}, we drop $z$ in the next equations for simplicity.
To make a distinction between the continuous spectrum and the discrete spectrum,
we denote the real frequencies by $\omega$ and the complex eigenvalues by $\lambda$.
\end{remark}
\begin{remark}
It is shown in \cite[P. 50, Eq. 6.23 ]{faddeev2007hamiltonian} that $a(\lambda)$ can be expressed in terms of
$b(\omega)$, $\omega\in\mathbb{R}$ and $N$ eigenvalues of the discrete part as 
\begin{equation}\label{eq:a_from_b}
a(\lambda) =\exp\{\frac{1}{2\pi j}\int_{-\infty}^{+\infty}\frac{\ln\left(1-|b(\omega)|^2\right)}{\omega-\lambda}\text{d}\omega\}\prod_{k=1}^N\frac{\lambda-\lambda_k}{\lambda-\lambda_k^*}
\end{equation}
when Im$(\lambda)>0$. As a result, $b(\omega)$ and $\{\lambda_k,b(\lambda_k)\}_{k=1}^N$ characterize the nonlinear Fourier spectrum completely.
\end{remark}
\subsection{Multi-Soliton Pulses}
An $N-$soliton pulse is only described by the discrete part which
contains $N$ pairs of eigenvalues and the b-coefficients, i.e. $\{\lambda_k,b(\lambda_k)\}_{k=1}^N$. Accordingly, 
\begin{equation}\label{eq:a_soliton}
a(\lambda) = \prod_{k=1}^N\frac{\lambda-\lambda_k}{\lambda-\lambda_k^*}.
\end{equation}

Darboux transformation (DT) is an effective algorithm to generate a multi-soliton pulse~\cite{matveev1991darboux}.
 It generates an $N-$soliton $q\left(t\right)$ recursively
by adding a pair $\{\lambda_k,b(\lambda_k)\}$ in each recursion.
 It can also be used to find the Jost solution of \eqref{eq:ZS} for any $\lambda$. For instance, one can show the following result on $b(\lambda)$.
 Let $\sigma_{\rm min}=\min_{k} \text{Im}(\lambda_k)$. It is known 
that the tails of the corresponding multi-soliton decays like $\exp(-2\sigma_{\rm min}|t|)$, e.g. see \cite{span2017timebandwidth}. 
Therefore, $b(\lambda)$ is analytic if  $-\sigma_{\rm min}<\text{Im}(\lambda)<+\sigma_{\rm min}$, 
see~\cite{ablowitz1981solitons}. For this range of $\lambda$, one can apply DT to find
$(v_1(t;z),v_2(t;z))$ and show that $v_2(t;z)= O(\exp(2(j\lambda-\sigma_{\rm min})t)$
when $t\to\infty$. Accordingly,
\begin{equation}
\label{eq:b_sol}
b(\lambda)=0, \text{ if }-\sigma_{\rm min}<\text{Im}(\lambda)<+\sigma_{\rm min}
\end{equation}
The same relation holds for $b^*(\lambda^*)$, too.
Now we introduce a large and practically interesting subset of multi-soliton pulses.
\begin{lemma}[Symmetric Multi-Solitons~\cite{span2017timebandwidth}] Let $q(t)$ be an $N-$soliton pulse of the discrete spectrum 
$\{\lambda_k,b(\lambda_k)\}_{k=1}^N$. Let $\lambda_k=j\sigma_k$ and $\sigma_k>0$ (pure imaginary eigenvalues). 
The pulse is symmetric, i.e. $q(t)=q(-t)$, if and only if $|b(\lambda_k)|=1$. 
\label{lem:symmetric_sol}
\end{lemma}

We discussed in \cite{span2017timebandwidth} that the symmetric multi-soliton have 
the smallest pulse duration (for the same set of eigenvalues) which can be of practical interest.

\section{Nonlinear Fourier Spectrum of Truncated Multi-Solitons}\label{sec:trun}

We outline the main results here and postpone their analysis to Section~\ref{sec:anal}. We present the results for the symmetric multi-solitons though the method is more general. This restriction allows obtaining  simple closed-form expressions.
Thus, we assume that $q(t)$ is an $N-$soliton pulse with the discrete spectrum
$\{\lambda_k,b(\lambda_k)\}_{k=1}^N$ such that $\lambda_k=j\sigma_k$ and $|b(\lambda_k)|=1$. Assume further 
that $\sigma_1=\min_{k}\sigma_k$ and define $\exp(j\phi)=b(\lambda_1)/|b(\lambda_1)|$. We define the truncated pulse  
as follows:
\begin{definition}[Truncation of a pulse] 
Let $q_{ \scriptscriptstyle T}(t)$ denote the truncated pulse of $q(t)$ defined as,
\begin{equation*}
q_{ \scriptscriptstyle T}(t)=
\begin{cases}
q(t), & |t|\leq T,\\
0, & |t|>T,
\end{cases}
\end{equation*}
We denote the Jost coefficients of $q_{ \scriptscriptstyle T}(t)$ by 
$(a_{ \scriptscriptstyle T}(\lambda),b_{ \scriptscriptstyle T}(\lambda))$, its continuous and discrete spectrum by
$\tilde{Q}_c(\omega)$ and $\{\tilde{\lambda}_k,\tilde{b}_k\}$.
\end{definition}

\subsection{Nonlinear Spectrum: $(a_{ \scriptscriptstyle T}(\lambda),b_{ \scriptscriptstyle T}(\lambda))$}\label{sec:con}

Now we present the following two theorems for the nonlinear spectrum of $q_{ \scriptscriptstyle T}(t)$. 
Both theorems are obtained based on approximation of the truncated tails. 
As we discuss in Section~\ref{sec:anal}, the approximation is rather accurate specially when $T$ is large.  
Let define $\alpha(\lambda)$ and $\beta(\lambda)$ for $\lambda\in \mathbb{C}$ as follows:
\begin{align}
\label{eq:a_trunc}
\alpha(\lambda) &= 1-\frac{e^{-2\sigma_1(T-t_0)}}{e^{-2\sigma_1(T-t_0)}+e^{2\sigma_1(T-t_0)}}
\left(\frac{2j\sigma_1}{\lambda+j\sigma_1}\right)\\
\beta(\lambda) &=e^{j\phi+jN\pi}\left(\frac{-2j\sigma_1}{\lambda+j\sigma_1}\right)
\frac{ e^{2j\lambda T}}{e^{-2\sigma_1(T-t_0)}+e^{2\sigma_1(T-t_0)}}
\label{eq:b_trunc}
\end{align}
where $\sigma_1=\min_{k}\sigma_k$, $\exp(j\phi)=b(\lambda_1)/|b(\lambda_1)|$, and
\begin{equation*}
t_0=\frac{1}{2\sigma_1}\sum_{k=2}^N \ln\left(\frac{\sigma_k+\sigma_1}{\sigma_k-\sigma_1}\right).
\end{equation*}

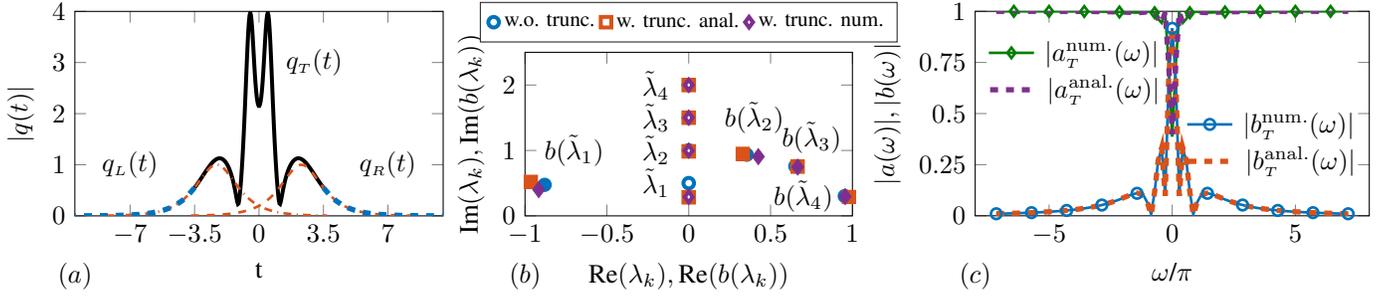
\begin{figure*}[t!]
\newlength{\wlength}
\newlength{\hlength}
\setlength{\wlength}{0.3\textwidth}
\setlength{\hlength}{0.15\textwidth}
\definecolor{mycolor1}{rgb}{0.00000,0.44700,0.74100}%
\definecolor{mycolor2}{rgb}{0.85000,0.32500,0.09800}%
\definecolor{mycolor3}{rgb}{0.49412,0.18431,0.55686}%
\definecolor{mycolor4}{rgb}{0.00000,0.49804,0.00000}%
\begin{tikzpicture}
\begin{scope}
\node[] at (0,-0.8) {$(a)$};
\begin{axis}[%
width=0.9\wlength,
height=\hlength,
at={(0\wlength,0\hlength)},
scale only axis,
xmin=-10,
xmax=10,
xtick={-7,-3.5,-0,3.5,7},
xlabel={t},
ymin=0,
ymax=4,
ytick={0, 1, 2, 3, 4},
ylabel={$|q(t)|$},
axis background/.style={fill=white},
legend style={legend cell align=left,align=left,draw=white!15!black}
]
\node at (3,3) {\normalsize $q_{ \scriptscriptstyle T}(t)$};
\node[align=center] at (7,1) {\normalsize $q_{ \scriptscriptstyle R}(t)$};
\node[align=center] at (-7,1) {\normalsize $q_{ \scriptscriptstyle L}(t)$};

\addplot [color=black,solid,line width=1.5pt,forget plot]
  table[row sep=crcr]{%
-3.5	0.550381332980692\\
-3.46482412060302	0.567089310764848\\
-3.42964824120603	0.584122020457039\\
-3.39447236180905	0.601467721558069\\
-3.35929648241206	0.619112572860806\\
-3.32412060301508	0.63704049141039\\
-3.28894472361809	0.655233013459908\\
-3.25376884422111	0.673669159476093\\
-3.21859296482412	0.692325305516822\\
-3.18341708542714	0.71117506356278\\
-3.14824120603015	0.730189173629022\\
-3.11306532663317	0.749335410696869\\
-3.07788944723618	0.768578509677653\\
-3.0427135678392	0.787880111732085\\
-3.00753768844221	0.807198735305235\\
-2.97236180904523	0.826489775179077\\
-2.93718592964824	0.845705532674033\\
-2.90201005025126	0.864795279831034\\
-2.86683417085427	0.883705359960449\\
-2.83165829145729	0.902379326341742\\
-2.7964824120603	0.92075812009081\\
-2.76130653266332	0.938780287278729\\
-2.72613065326633	0.95638223429371\\
-2.69095477386935	0.973498519201169\\
-2.65577889447236	0.990062175502192\\
-2.62060301507538	1.00600506325204\\
-2.58542713567839	1.02125824102436\\
-2.55025125628141	1.03575235074734\\
-2.51507537688442	1.04941800605686\\
-2.47989949748744	1.06218617357343\\
-2.44472361809045	1.07398853548125\\
-2.40954773869347	1.084757821029\\
-2.37437185929648	1.09442809413711\\
-2.3391959798995	1.10293498422183\\
-2.30402010050251	1.11021584765406\\
-2.26884422110553	1.11620984795714\\
-2.23366834170854	1.12085794388622\\
-2.19849246231156	1.12410277587083\\
-2.16331658291457	1.12588844286686\\
-2.12814070351759	1.12616016335928\\
-2.0929648241206	1.12486381597649\\
-2.05778894472362	1.12194535680811\\
-2.02261306532663	1.11735011195664\\
-1.98743718592965	1.1110219450146\\
-1.95226130653266	1.10290229999366\\
-1.91708542713568	1.09292912074814\\
-1.88190954773869	1.08103564822233\\
-1.84673366834171	1.06714909711705\\
-1.81155778894472	1.05118921420378\\
-1.77638190954774	1.03306672217263\\
-1.74120603015075	1.01268165667998\\
-1.70603015075377	0.989921611982933\\
-1.67085427135678	0.964659925283094\\
-1.6356783919598	0.936753856956442\\
-1.60050251256281	0.906042872634074\\
-1.56532663316583	0.872347220874927\\
-1.53015075376884	0.835467159666011\\
-1.49497487437186	0.79518348076347\\
-1.45979899497487	0.75126054597182\\
-1.42462311557789	0.703454171740361\\
-1.3894472361809	0.651529036552412\\
-1.35427135678392	0.595295443239339\\
-1.31909547738693	0.53468742939266\\
-1.28391959798995	0.469935117605057\\
-1.24874371859296	0.401968902260703\\
-1.21356783919598	0.333437033216434\\
-1.17839195979899	0.27135114876883\\
-1.14321608040201	0.232595444988754\\
-1.10804020100502	0.243430934503803\\
-1.07286432160804	0.312524040577936\\
-1.03768844221106	0.424569964201047\\
-1.00251256281407	0.566400735748274\\
-0.967336683417086	0.732450344808394\\
-0.9321608040201	0.920992408553071\\
-0.896984924623116	1.13176251980063\\
-0.86180904522613	1.36481075003453\\
-0.826633165829146	1.61979452821221\\
-0.791457286432161	1.895367845653\\
-0.756281407035176	2.18854499827692\\
-0.721105527638191	2.49403240189898\\
-0.685929648241206	2.80361901613603\\
-0.650753768844221	3.1058200465282\\
-0.615577889447236	3.38605965119912\\
-0.580402010050251	3.62769226457842\\
-0.545226130653266	3.81401157204555\\
-0.510050251256281	3.93104010766678\\
-0.474874371859296	3.97043466066599\\
-0.439698492462312	3.93155452276122\\
-0.404522613065327	3.82189660694499\\
-0.369346733668342	3.65571150681623\\
-0.334170854271357	3.45133169865648\\
-0.298994974874372	3.22813227640168\\
-0.263819095477387	3.00394406834018\\
-0.228643216080402	2.7933388471345\\
-0.193467336683417	2.60681828264197\\
-0.158291457286432	2.45074441856139\\
-0.123115577889447	2.3278336231733\\
-0.0879396984924625	2.23807182726279\\
-0.0527638190954773	2.17989336598708\\
-0.0175879396984926	2.15140080765904\\
0.0175879396984926	2.15136523649234\\
0.0527638190954773	2.17978663831913\\
0.0879396984924625	2.23789421441044\\
0.123115577889447	2.32758653974288\\
0.158291457286432	2.45043160449567\\
0.193467336683417	2.60644704198679\\
0.228643216080402	2.79292112646122\\
0.263819095477387	3.00349722405479\\
0.298994974874372	3.22767934703003\\
0.334170854271357	3.45090085530194\\
0.369346733668342	3.65533425300353\\
0.404522613065327	3.8216044593682\\
0.439698492462312	3.93137444537339\\
0.474874371859296	3.97038428794843\\
0.510050251256281	3.93112431302863\\
0.545226130653266	3.81422171879268\\
0.580402010050252	3.62800847005162\\
0.615577889447236	3.38645517144532\\
0.650753768844221	3.10626620500971\\
0.685929648241206	2.80408926024155\\
0.721105527638191	2.49450477385984\\
0.756281407035176	2.1890030301128\\
0.791457286432161	1.8958003064904\\
0.826633165829146	1.62019454012382\\
0.86180904522613	1.36517467784157\\
0.896984924623116	1.13208886739836\\
0.932160804020101	0.921280790537702\\
0.967336683417085	0.732700448454699\\
1.00251256281407	0.566610977634937\\
1.03768844221106	0.424735142346701\\
1.07286432160804	0.312631013713025\\
1.10804020100502	0.243457523882469\\
1.14321608040201	0.232540231531358\\
1.178391959799	0.271254574181386\\
1.21356783919598	0.333332166684032\\
1.24874371859296	0.401868909680326\\
1.28391959798995	0.469844197420116\\
1.31909547738693	0.534606598893053\\
1.35427135678392	0.595224583940219\\
1.3894472361809	0.651467618619984\\
1.42462311557789	0.703401527763148\\
1.45979899497487	0.751215979779987\\
1.49497487437186	0.795146311007919\\
1.53015075376884	0.835436736000002\\
1.56532663316583	0.872322928259604\\
1.60050251256281	0.906024130555089\\
1.6356783919598	0.936740116664281\\
1.67085427135678	0.964650666524428\\
1.70603015075377	0.989916339914039\\
1.74120603015075	1.01267989927077\\
1.77638190954774	1.03306802821488\\
1.81155778894472	1.05119315193531\\
1.84673366834171	1.06715525341934\\
1.88190954773869	1.0810436283212\\
1.91708542713568	1.09293854833261\\
1.95226130653266	1.10291281766292\\
1.98743718592965	1.11103321496214\\
2.02261306532663	1.11736181680373\\
2.05778894472362	1.12195720050418\\
2.0929648241206	1.12487552469108\\
2.12814070351759	1.12617148629911\\
2.16331658291457	1.12589915296118\\
2.19849246231156	1.12411267027876\\
2.23366834170854	1.12086684429057\\
2.26884422110553	1.11621760062064\\
2.30402010050251	1.11022232322403\\
2.3391959798995	1.10294007728032\\
2.37437185929648	1.09443172250225\\
2.40954773869347	1.08475992482214\\
2.44472361809045	1.07398907598121\\
2.47989949748744	1.06218513188488\\
2.51507537688442	1.04941538162396\\
2.55025125628141	1.03574815974652\\
2.58542713567839	1.02125251467105\\
2.62060301507538	1.00599784605646\\
2.65577889447236	0.990053523507564\\
2.69095477386935	0.973488498235784\\
2.72613065326633	0.956370918265104\\
2.76130653266332	0.938767756534531\\
2.7964824120603	0.920744459867046\\
2.83165829145729	0.902364625315074\\
2.86683417085427	0.883689708916763\\
2.90201005025126	0.864778770458946\\
2.93718592964824	0.845688256487994\\
2.97236180904523	0.826471822573595\\
3.00753768844221	0.80718019473883\\
3.0427135678392	0.787861069037036\\
3.07788944723618	0.7685590474897\\
3.11306532663317	0.749315607997486\\
3.14824120603015	0.73016910539176\\
3.18341708542714	0.711154800494398\\
3.21859296482412	0.692304913883556\\
3.25376884422111	0.673648701005371\\
3.28894472361809	0.655212545307874\\
3.32412060301507	0.637020066185952\\
3.35929648241206	0.619092238697466\\
3.39447236180905	0.601447522225291\\
3.42964824120603	0.584101995503644\\
3.46482412060301	0.567069495687643\\
3.5	0.550361759412155\\
};
\addplot [color=mycolor2,dashed,line width=1.0pt,forget plot]
  table[row sep=crcr]{%
-3.69743690748997	0.00495747389356038\\
-3.59743690748997	0.0054788485212007\\
-3.49743690748997	0.00605505398959552\\
-3.39743690748997	0.00669185599637017\\
-3.29743690748997	0.00739562630372176\\
-3.19743690748997	0.00817340636714055\\
-3.09743690748997	0.00903297761696762\\
-2.99743690748997	0.0099829390788132\\
-2.89743690748997	0.0110327930862334\\
-2.79743690748997	0.012193039911852\\
-2.69743690748997	0.0134752822213046\\
-2.59743690748997	0.0148923403377577\\
-2.49743690748997	0.0164583793927991\\
-2.39743690748997	0.0181890495312658\\
-2.29743690748997	0.020101640431542\\
-2.19743690748997	0.0222152514966497\\
-2.09743690748997	0.0245509791615513\\
-1.99743690748997	0.0271321228433919\\
-1.89743690748997	0.0299844111266583\\
-1.79743690748997	0.0331362498142155\\
-1.69743690748997	0.0366189934736865\\
-1.59743690748997	0.0404672420470393\\
-1.49743690748997	0.0447191639426338\\
-1.39743690748997	0.0494168467566525\\
-1.29743690748997	0.0546066763249983\\
-1.19743690748997	0.0603397441201676\\
-1.09743690748997	0.0666722819899217\\
-0.997436907489968	0.0736661217649808\\
-0.897436907489968	0.0813891751807532\\
-0.797436907489968	0.0899159266508794\\
-0.697436907489969	0.0993279274194332\\
-0.597436907489969	0.109714274141502\\
-0.497436907489968	0.121172047532414\\
-0.397436907489968	0.133806676793102\\
-0.297436907489968	0.147732182327837\\
-0.197436907489969	0.163071231929978\\
-0.0974369074899686	0.179954923081637\\
0.00256309251003151	0.198522175149339\\
0.102563092510032	0.218918578920168\\
0.202563092510032	0.241294506201855\\
0.302563092510031	0.26580222883408\\
0.402563092510032	0.292591735483763\\
0.502563092510032	0.321804869506588\\
0.602563092510031	0.353567349501402\\
0.702563092510032	0.38797818987449\\
0.802563092510031	0.42509603494228\\
0.902563092510032	0.464921992408982\\
1.00256309251003	0.507378750740602\\
1.10256309251003	0.552286154278205\\
1.20256309251003	0.599334060570793\\
1.30256309251003	0.648054273663885\\
1.40256309251003	0.697794641100333\\
1.50256309251003	0.74769991823742\\
1.60256309251003	0.796705459992875\\
1.70256309251003	0.843550687621807\\
1.80256309251003	0.886818883970074\\
1.90256309251003	0.925007451905755\\
2.00256309251003	0.956627911900248\\
2.10256309251003	0.980327997644726\\
2.20256309251003	0.995020748953227\\
2.30256309251003	1\\
2.40256309251003	0.995020748953227\\
2.50256309251003	0.980327997644726\\
2.60256309251003	0.956627911900248\\
2.70256309251003	0.925007451905755\\
2.80256309251003	0.886818883970074\\
2.90256309251003	0.843550687621807\\
3.00256309251003	0.796705459992875\\
3.10256309251003	0.74769991823742\\
3.20256309251003	0.697794641100333\\
3.30256309251003	0.648054273663885\\
3.40256309251003	0.599334060570793\\
3.50256309251003	0.552286154278205\\
3.60256309251003	0.507378750740602\\
3.70256309251003	0.464921992408982\\
3.80256309251003	0.42509603494228\\
3.90256309251003	0.38797818987449\\
4.00256309251003	0.353567349501402\\
4.10256309251003	0.321804869506588\\
4.20256309251003	0.292591735483763\\
4.30256309251003	0.26580222883408\\
4.40256309251003	0.241294506201855\\
4.50256309251003	0.218918578920168\\
4.60256309251003	0.198522175149339\\
4.70256309251003	0.179954923081637\\
4.80256309251003	0.163071231929978\\
4.90256309251003	0.147732182327837\\
5.00256309251003	0.133806676793102\\
5.10256309251003	0.121172047532414\\
5.20256309251003	0.109714274141502\\
5.30256309251003	0.0993279274194332\\
5.40256309251003	0.0899159266508794\\
5.50256309251003	0.0813891751807532\\
5.60256309251003	0.0736661217649808\\
5.70256309251003	0.0666722819899217\\
5.80256309251003	0.0603397441201676\\
5.90256309251003	0.0546066763249983\\
6.00256309251003	0.0494168467566525\\
6.10256309251003	0.0447191639426338\\
6.20256309251003	0.0404672420470393\\
6.30256309251003	0.0366189934736865\\
6.40256309251003	0.0331362498142155\\
6.50256309251003	0.0299844111266583\\
6.60256309251003	0.0271321228433919\\
6.70256309251003	0.0245509791615513\\
6.80256309251003	0.0222152514966497\\
6.90256309251003	0.020101640431542\\
7.00256309251003	0.0181890495312658\\
7.10256309251003	0.0164583793927991\\
7.20256309251003	0.0148923403377577\\
7.30256309251003	0.0134752822213046\\
7.40256309251003	0.012193039911852\\
7.50256309251003	0.0110327930862334\\
7.60256309251003	0.0099829390788132\\
7.70256309251003	0.00903297761696762\\
7.80256309251003	0.00817340636714055\\
7.90256309251003	0.00739562630372176\\
8.00256309251003	0.00669185599637017\\
8.10256309251003	0.00605505398959552\\
8.20256309251003	0.0054788485212007\\
8.30256309251003	0.00495747389356038\\
};
\addplot [color=mycolor2,dashdotted,line width=1.0pt,forget plot]
  table[row sep=crcr]{%
-8.30256309251003	0.00495747389356038\\
-8.20256309251003	0.0054788485212007\\
-8.10256309251003	0.00605505398959552\\
-8.00256309251003	0.00669185599637017\\
-7.90256309251003	0.00739562630372176\\
-7.80256309251003	0.00817340636714055\\
-7.70256309251003	0.00903297761696762\\
-7.60256309251003	0.0099829390788132\\
-7.50256309251003	0.0110327930862334\\
-7.40256309251003	0.012193039911852\\
-7.30256309251003	0.0134752822213046\\
-7.20256309251003	0.0148923403377577\\
-7.10256309251003	0.0164583793927991\\
-7.00256309251003	0.0181890495312658\\
-6.90256309251003	0.020101640431542\\
-6.80256309251003	0.0222152514966497\\
-6.70256309251003	0.0245509791615513\\
-6.60256309251003	0.0271321228433919\\
-6.50256309251003	0.0299844111266583\\
-6.40256309251003	0.0331362498142155\\
-6.30256309251003	0.0366189934736865\\
-6.20256309251003	0.0404672420470393\\
-6.10256309251003	0.0447191639426338\\
-6.00256309251003	0.0494168467566525\\
-5.90256309251003	0.0546066763249983\\
-5.80256309251003	0.0603397441201676\\
-5.70256309251003	0.0666722819899217\\
-5.60256309251003	0.0736661217649808\\
-5.50256309251003	0.0813891751807532\\
-5.40256309251003	0.0899159266508794\\
-5.30256309251003	0.0993279274194332\\
-5.20256309251003	0.109714274141502\\
-5.10256309251003	0.121172047532414\\
-5.00256309251003	0.133806676793102\\
-4.90256309251003	0.147732182327837\\
-4.80256309251003	0.163071231929978\\
-4.70256309251003	0.179954923081637\\
-4.60256309251003	0.198522175149339\\
-4.50256309251003	0.218918578920168\\
-4.40256309251003	0.241294506201855\\
-4.30256309251003	0.26580222883408\\
-4.20256309251003	0.292591735483763\\
-4.10256309251003	0.321804869506588\\
-4.00256309251003	0.353567349501402\\
-3.90256309251003	0.38797818987449\\
-3.80256309251003	0.42509603494228\\
-3.70256309251003	0.464921992408982\\
-3.60256309251003	0.507378750740602\\
-3.50256309251003	0.552286154278205\\
-3.40256309251003	0.599334060570793\\
-3.30256309251003	0.648054273663885\\
-3.20256309251003	0.697794641100333\\
-3.10256309251003	0.74769991823742\\
-3.00256309251003	0.796705459992875\\
-2.90256309251003	0.843550687621807\\
-2.80256309251003	0.886818883970074\\
-2.70256309251003	0.925007451905755\\
-2.60256309251003	0.956627911900248\\
-2.50256309251003	0.980327997644726\\
-2.40256309251003	0.995020748953227\\
-2.30256309251003	1\\
-2.20256309251003	0.995020748953227\\
-2.10256309251003	0.980327997644726\\
-2.00256309251003	0.956627911900248\\
-1.90256309251003	0.925007451905755\\
-1.80256309251003	0.886818883970074\\
-1.70256309251003	0.843550687621807\\
-1.60256309251003	0.796705459992875\\
-1.50256309251003	0.74769991823742\\
-1.40256309251003	0.697794641100333\\
-1.30256309251003	0.648054273663885\\
-1.20256309251003	0.599334060570793\\
-1.10256309251003	0.552286154278205\\
-1.00256309251003	0.507378750740602\\
-0.902563092510032	0.464921992408982\\
-0.802563092510031	0.42509603494228\\
-0.702563092510032	0.38797818987449\\
-0.602563092510031	0.353567349501402\\
-0.502563092510032	0.321804869506588\\
-0.402563092510032	0.292591735483763\\
-0.302563092510031	0.26580222883408\\
-0.202563092510032	0.241294506201855\\
-0.102563092510032	0.218918578920168\\
-0.00256309251003151	0.198522175149339\\
0.0974369074899686	0.179954923081637\\
0.197436907489969	0.163071231929978\\
0.297436907489968	0.147732182327837\\
0.397436907489968	0.133806676793102\\
0.497436907489968	0.121172047532414\\
0.597436907489969	0.109714274141502\\
0.697436907489969	0.0993279274194332\\
0.797436907489968	0.0899159266508794\\
0.897436907489968	0.0813891751807532\\
0.997436907489968	0.0736661217649808\\
1.09743690748997	0.0666722819899217\\
1.19743690748997	0.0603397441201676\\
1.29743690748997	0.0546066763249983\\
1.39743690748997	0.0494168467566525\\
1.49743690748997	0.0447191639426338\\
1.59743690748997	0.0404672420470393\\
1.69743690748997	0.0366189934736865\\
1.79743690748997	0.0331362498142155\\
1.89743690748997	0.0299844111266583\\
1.99743690748997	0.0271321228433919\\
2.09743690748997	0.0245509791615513\\
2.19743690748997	0.0222152514966497\\
2.29743690748997	0.020101640431542\\
2.39743690748997	0.0181890495312658\\
2.49743690748997	0.0164583793927991\\
2.59743690748997	0.0148923403377577\\
2.69743690748997	0.0134752822213046\\
2.79743690748997	0.012193039911852\\
2.89743690748997	0.0110327930862334\\
2.99743690748997	0.0099829390788132\\
3.09743690748997	0.00903297761696762\\
3.19743690748997	0.00817340636714055\\
3.29743690748997	0.00739562630372176\\
3.39743690748997	0.00669185599637017\\
3.49743690748997	0.00605505398959552\\
3.59743690748997	0.0054788485212007\\
3.69743690748997	0.00495747389356038\\
};
\addplot [color=mycolor1,dashed,line width=2.0pt,forget plot]
  table[row sep=crcr]{%
-3.5	0.550381332980692\\
-3.84210526315789	0.406099392785048\\
-4.18421052631579	0.294790417381619\\
-4.52631578947368	0.212039613800599\\
-4.86842105263158	0.151735372611161\\
-5.21052631578947	0.108264111400156\\
-5.55263157894737	0.0771154935918261\\
-5.89473684210526	0.0548726814014381\\
-6.23684210526316	0.0390210322148673\\
-6.57894736842105	0.0277376470867927\\
-6.92105263157895	0.01971194037325\\
-7.26315789473684	0.0140060509032141\\
-7.60526315789474	0.0099506770591898\\
-7.94736842105263	0.00706896575037725\\
-8.28947368421053	0.00502152814965935\\
-8.63157894736842	0.00356697271712496\\
-8.97368421052632	0.00253368365815585\\
-9.31578947368421	0.00179968737235996\\
-9.65789473684211	0.00127830994068818\\
-10	0.000907969627211078\\
};
\addplot [color=mycolor1,dashed,line width=2.0pt,forget plot]
  table[row sep=crcr]{%
3.5	0.550361759412155\\
3.84210526315789	0.406083165791443\\
4.18421052631579	0.294777966864661\\
4.52631578947368	0.212030423805262\\
4.86842105263158	0.151728720144158\\
5.21052631578947	0.108259342571806\\
5.55263157894737	0.0771120915795937\\
5.89473684210526	0.0548702602135481\\
6.23684210526316	0.0390193110396312\\
6.57894736842105	0.0277364241896198\\
6.92105263157895	0.0197110717099397\\
7.26315789473684	0.0140054339237213\\
7.60526315789474	0.00995023885583653\\
7.94736842105263	0.00706865452250043\\
8.28947368421053	0.00502130710309477\\
8.63157894736842	0.00356681571958347\\
8.97368421052632	0.00253357215024925\\
9.31578947368421	0.0017996081730055\\
9.65789473684211	0.00127825368840777\\
10	0.000907929673185705\\
};
\end{axis}
\end{scope}
\begin{scope}[xshift=1.1\wlength]
\node[] at (0,-0.8) {$(b)$};
\begin{axis}[%
width=0.8\wlength,
height=0.8\hlength,
at={(0\wlength,0\hlength)},
scale only axis,
xmin=-1,
xmax=1,
xtick={  -1, -0.5,    0,  0.5,    1},
xlabel={$\text{Re}(\lambda_k), \text{Re}(b(\lambda_k))$},
ymin=0,
ymax=2.5,
ylabel={$\text{Im}(\lambda_k), \text{Im}(b(\lambda_k))$},
axis background/.style={fill=white},
legend style={legend cell align=left,align=left,draw=white!15!black,at={(axis cs:0.,2.65)},anchor=south,legend columns=3,font=\footnotesize}
]
\node at (-0.2,0.5) {\normalsize $\tilde{\lambda}_1$};
\node at (-0.2,1.) {\normalsize $\tilde{\lambda}_2$};
\node at (-0.2,1.5) {\normalsize $\tilde{\lambda}_3$};
\node at (-0.2,2.) {\normalsize $\tilde{\lambda}_4$};
\node at (-0.7,1) {\normalsize $b(\tilde{\lambda}_1)$};
\node at (0.4,1.5) {\normalsize $b(\tilde{\lambda}_2)$};
\node at (0.75,1.2) {\normalsize $b(\tilde{\lambda}_3)$};
\node at (0.7,0.3) {\normalsize $b(\tilde{\lambda}_4)$};

\addplot [color=mycolor1,line width=1.5pt,only marks,mark=o,mark options={solid}]
  table[row sep=crcr]{%
0	0.5\\
0	1\\
0	1.5\\
0	2\\
};
\addlegendentry{w.o. trunc.};

\addplot [color=mycolor2,line width=1.5pt,only marks,mark=square,mark options={solid}]
  table[row sep=crcr]{%
0	0.284000710031204\\
0	0.986189576843832\\
0	1.50073468734318\\
0	1.99998719380006\\
};
\addlegendentry{w. trunc. anal.};

\addplot [color=mycolor3,line width=1.5pt,only marks,mark=diamond,mark options={solid}]
  table[row sep=crcr]{%
0	0.288672262307018\\
0	0.997378146157312\\
0	1.50061243094074\\
0	1.99998302594938\\
};
\addlegendentry{w. trunc. num.};

\addplot [color=mycolor1,mark size=2.5pt,only marks,mark=*,mark options={solid,fill=mycolor1}]
  table[row sep=crcr]{%
-0.880230501384588	0.474500015207836\\
0.354548125817861	0.935051338953665\\
0.651011270888135	0.759132910086643\\
0.953280505077772	0.302264219911434\\
};

\addplot [color=mycolor2,mark size=2.5pt,only marks,mark=square*,mark options={solid,fill=mycolor2}]
  table[row sep=crcr]{%
-0.964528997683929	0.519942246206554\\
0.32999482992804	0.948290149980413\\
0.665343960062426	0.751410862075868\\
0.97829225782782	0.288783089403638\\
};

\addplot [color=mycolor3,mark size=3.0pt,only marks,mark=diamond*,mark options={solid,fill=mycolor3}]
  table[row sep=crcr]{%
-0.916251131455782	0.400587172745122\\
0.425057608650303	0.905192272667699\\
0.66577918251607	0.746222970952488\\
0.955601198308601	0.294859309012663\\
};

\end{axis}
\end{scope}
\begin{scope}[xshift=2.2\wlength]
\node[] at (0,-0.8) {$(c)$};
\begin{axis}[%
width=0.963\wlength,
height=\hlength,
at={(0\wlength,0\hlength)},
scale only axis,
xmin=-8,
xmax=8,
xlabel={$\omega\text{/}\pi$},
ymin=0,
ymax=1.0,
ytick={   0, 0.25,  0.5, 0.75,    1},
ylabel={$|a(\omega)|,|b(\omega)|$},
axis background/.style={fill=white},
]
\addplot [color=mycolor4,solid,line width=1.0pt,mark=diamond,mark repeat=10,mark phase=6,mark options={solid,fill=mycolor3}]
  table[row sep=crcr]{%
-7.14285714285714	0.999949707375009\\
-7	0.999945616037407\\
-6.85714285714286	0.999941103487991\\
-6.71428571428571	0.99993611707562\\
-6.57142857142857	0.999930596387448\\
-6.42857142857143	0.999924471928763\\
-6.28571428571429	0.999917663548284\\
-6.14285714285714	0.999910078554122\\
-6	0.999901609452617\\
-5.85714285714286	0.99989213122595\\
-5.71428571428571	0.99988149804385\\
-5.57142857142857	0.999869539278709\\
-5.42857142857143	0.999856054660521\\
-5.28571428571429	0.999840808366225\\
-5.14285714285714	0.99982352178497\\
-5	0.999803864633238\\
-4.85714285714286	0.999781444007848\\
-4.71428571428571	0.999755790855748\\
-4.57142857142857	0.999726343201292\\
-4.42857142857143	0.999692425297758\\
-4.28571428571429	0.999653221652953\\
-4.14285714285714	0.999607744612509\\
-4	0.999554793866163\\
-3.85714285714286	0.99949290587827\\
-3.71428571428571	0.999420290861334\\
-3.57142857142857	0.999334754580901\\
-3.42857142857143	0.999233602158731\\
-3.28571428571429	0.999113521456026\\
-3.14285714285714	0.998970445226281\\
-3	0.99879939531953\\
-2.85714285714286	0.99859432130436\\
-2.71428571428571	0.998347964783323\\
-2.57142857142857	0.998051818658548\\
-2.42857142857143	0.997696325202338\\
-2.28571428571429	0.997271600794698\\
-2.14285714285714	0.996769248115263\\
-2	0.996186319849334\\
-1.85714285714286	0.995533379786165\\
-1.71428571428571	0.994849986385351\\
-1.57142857142857	0.994232449373097\\
-1.42857142857143	0.993877733878233\\
-1.28571428571429	0.994132417955234\\
-1.14285714285714	0.995465981592449\\
-1	0.998046749646464\\
-0.857142857142857	0.999994079213197\\
-0.714285714285714	0.993054228236899\\
-0.571428571428571	0.962625802811508\\
-0.428571428571429	0.943360102030744\\
-0.285714285714286	0.995147910941101\\
-0.142857142857143	0.737631310225877\\
0	0.387348672356202\\
0.142857142857143	0.737631310225877\\
0.285714285714286	0.995147910941101\\
0.428571428571429	0.943360102030744\\
0.571428571428571	0.962625802811508\\
0.714285714285714	0.993054228236899\\
0.857142857142857	0.999994079213197\\
1	0.998046749646464\\
1.14285714285714	0.995465981592449\\
1.28571428571429	0.994132417955234\\
1.42857142857143	0.993877733878233\\
1.57142857142857	0.994232449373097\\
1.71428571428571	0.994849986385351\\
1.85714285714286	0.995533379786165\\
2	0.996186319849334\\
2.14285714285714	0.996769248115263\\
2.28571428571429	0.997271600794698\\
2.42857142857143	0.997696325202338\\
2.57142857142857	0.998051818658548\\
2.71428571428571	0.998347964783323\\
2.85714285714286	0.99859432130436\\
3	0.99879939531953\\
3.14285714285714	0.998970445226281\\
3.28571428571429	0.999113521456026\\
3.42857142857143	0.999233602158731\\
3.57142857142857	0.999334754580901\\
3.71428571428571	0.999420290861334\\
3.85714285714286	0.99949290587827\\
4	0.999554793866163\\
4.14285714285714	0.999607744612509\\
4.28571428571429	0.999653221652953\\
4.42857142857143	0.999692425297758\\
4.57142857142857	0.999726343201292\\
4.71428571428571	0.999755790855748\\
4.85714285714286	0.999781444007848\\
5	0.999803864633238\\
5.14285714285714	0.99982352178497\\
5.28571428571429	0.999840808366225\\
5.42857142857143	0.999856054660521\\
5.57142857142857	0.999869539278709\\
5.71428571428571	0.99988149804385\\
5.85714285714286	0.99989213122595\\
6	0.999901609452617\\
6.14285714285714	0.999910078554122\\
6.28571428571429	0.999917663548284\\
6.42857142857143	0.999924471928763\\
6.57142857142857	0.999930596387448\\
6.71428571428571	0.99993611707562\\
6.85714285714286	0.999941103487991\\
7	0.999945616037407\\
7.14285714285714	0.999949707375009\\
};
\label{p1}
\addplot [color=mycolor1,solid,line width=1.0pt,mark=o,mark repeat=10,mark options={solid,fill=mycolor1}]
  table[row sep=crcr]{%
-7.14285714285714	0.00996421447738084\\
-7	0.010361659750894\\
-6.85714285714286	0.0107830497522055\\
-6.71428571428571	0.0112303181818685\\
-6.57142857142857	0.0117055945811449\\
-6.42857142857143	0.0122112276657496\\
-6.28571428571429	0.0127498131280939\\
-6.14285714285714	0.013324223470629\\
-6	0.0139376447091157\\
-5.85714285714286	0.0145936156761137\\
-5.71428571428571	0.015296075308344\\
-5.57142857142857	0.016049416817475\\
-5.42857142857143	0.0168585467292806\\
-5.28571428571429	0.0177289616492338\\
-5.14285714285714	0.0186668217743184\\
-5	0.0196790572577472\\
-4.85714285714286	0.0207734660810385\\
-4.71428571428571	0.021958850038734\\
-4.57142857142857	0.0232451608624381\\
-4.42857142857143	0.0246436544591105\\
-4.28571428571429	0.026167121499376\\
-4.14285714285714	0.0278300461491031\\
-4	0.0296489439825566\\
-3.85714285714286	0.0316425400532444\\
-3.71428571428571	0.0338321665690451\\
-3.57142857142857	0.0362420369751165\\
-3.42857142857143	0.0388995185285344\\
-3.28571428571429	0.0418355967566994\\
-3.14285714285714	0.0450846688425368\\
-3	0.0486850414057332\\
-2.85714285714286	0.052677692124975\\
-2.71428571428571	0.0571057011432923\\
-2.57142857142857	0.0620108184048792\\
-2.42857142857143	0.067428310151174\\
-2.28571428571429	0.0733774526251203\\
-2.14285714285714	0.0798421986606856\\
-2	0.0867415046824155\\
-1.85714285714286	0.093868523635246\\
-1.71428571428571	0.100792614255077\\
-1.57142857142857	0.106672244154892\\
-1.42857142857143	0.109934268438331\\
-1.28571428571429	0.107702689112268\\
-1.14285714285714	0.0948553967295606\\
-1	0.0627018480940144\\
-0.857142857142857	0.00872519840116734\\
-0.714285714285714	0.116791405491516\\
-0.571428571428571	0.268758320374367\\
-0.428571428571429	0.329859295008875\\
-0.285714285714286	0.103530780070303\\
-0.142857142857143	0.669714323995102\\
0	0.916983408930505\\
0.142857142857143	0.669717585097421\\
0.285714285714286	0.103531468151343\\
0.428571428571429	0.329861639388943\\
0.571428571428571	0.268756436703805\\
0.714285714285714	0.11678875985561\\
0.857142857142857	0.00873092517452223\\
1	0.0627048697915733\\
1.14285714285714	0.0948573878356067\\
1.28571428571429	0.107703995249776\\
1.42857142857143	0.109935074015796\\
1.57142857142857	0.106672682185232\\
1.71428571428571	0.100792780725081\\
1.85714285714286	0.0938684916351833\\
2	0.0867413256841073\\
2.14285714285714	0.0798419126249372\\
2.28571428571429	0.0733770878153384\\
2.42857142857143	0.0674278882892522\\
2.57142857142857	0.0620103557055008\\
2.71428571428571	0.0571052094042528\\
2.85714285714286	0.0526771809262346\\
3	0.0486845172903403\\
3.14285714285714	0.0450841374336378\\
3.28571428571429	0.0418350618619821\\
3.42857142857143	0.0388989833311962\\
3.57142857142857	0.0362415037942065\\
3.71428571428571	0.0338316371316808\\
3.85714285714286	0.0316420157705516\\
4	0.0296484257667987\\
4.14285714285714	0.0278295348020024\\
4.28571428571429	0.0261666174943056\\
4.42857142857143	0.0246431581875246\\
4.57142857142857	0.0232446725508757\\
4.71428571428571	0.0219583698212886\\
4.85714285714286	0.02077299402879\\
5	0.0196785933575959\\
5.14285714285714	0.0186663659915945\\
5.28571428571429	0.0177285138921412\\
5.42857142857143	0.0168581068930581\\
5.57142857142857	0.0160489847686247\\
5.71428571428571	0.0152956509004954\\
5.85714285714286	0.0145931987529437\\
6	0.0139372351035928\\
6.14285714285714	0.0133238210143363\\
6.28571428571429	0.0127494176467061\\
6.42857142857143	0.0122108389861844\\
6.57142857142857	0.0117052125292757\\
6.71428571428571	0.0112299425854914\\
6.85714285714286	0.0107826804413327\\
7	0.0103612965580344\\
7.14285714285714	0.00996385723880991\\
};
\label{p2}
\addplot [color=mycolor2,dashed,line width=2.0pt,forget plot]
  table[row sep=crcr]{%
-7.14285714285714	0.0100290937095453\\
-7	0.0104290444226995\\
-6.85714285714286	0.0108530896623543\\
-6.71428571428571	0.0113031751172695\\
-6.57142857142857	0.0117814433853187\\
-6.42857142857143	0.0122902578485364\\
-6.28571428571429	0.0128322298974175\\
-6.14285714285714	0.0134102500308495\\
-6	0.0140275234473798\\
-5.85714285714286	0.0146876108481638\\
-5.71428571428571	0.0153944752943099\\
-5.57142857142857	0.0161525361037253\\
-5.42857142857143	0.0169667309372511\\
-5.28571428571429	0.0178425874125187\\
-5.14285714285714	0.0187863057970114\\
-5	0.0198048545675675\\
-4.85714285714286	0.0209060808757388\\
-4.71428571428571	0.0220988382137671\\
-4.57142857142857	0.0233931338083026\\
-4.42857142857143	0.0248002984313796\\
-4.28571428571429	0.0263331813283445\\
-4.14285714285714	0.0280063726800398\\
-4	0.029836455204528\\
-3.85714285714286	0.0318422847643186\\
-3.71428571428571	0.034045296512829\\
-3.57142857142857	0.0364698270729367\\
-3.42857142857143	0.0391434326150134\\
-3.28571428571429	0.0420971643194595\\
-3.14285714285714	0.0453657311679928\\
-3	0.0489874260330385\\
-2.85714285714286	0.0530035985446782\\
-2.71428571428571	0.0574572990402141\\
-2.57142857142857	0.0623904421555522\\
-2.42857142857143	0.0678383569800374\\
-2.28571428571429	0.0738197415897614\\
-2.14285714285714	0.0803185284460129\\
-2	0.0872514535411281\\
-1.85714285714286	0.094410220482399\\
-1.71428571428571	0.101358298077004\\
-1.57142857142857	0.107246615860701\\
-1.42857142857143	0.110485519870562\\
-1.28571428571429	0.108169938386228\\
-1.14285714285714	0.0951182395347092\\
-1	0.0624714776528369\\
-0.857142857142857	0.0034411536655177\\
-0.714285714285714	0.117657553012195\\
-0.571428571428571	0.270834938221604\\
-0.428571428571429	0.331770580215522\\
-0.285714285714286	0.0983902197861305\\
-0.142857142857143	0.675203710130844\\
0	0.921933298034021\\
0.142857142857143	0.675203710130844\\
0.285714285714286	0.0983902197861304\\
0.428571428571429	0.331770580215522\\
0.571428571428571	0.270834938221604\\
0.714285714285714	0.117657553012195\\
0.857142857142857	0.00344115366551768\\
1	0.0624714776528369\\
1.14285714285714	0.0951182395347092\\
1.28571428571429	0.108169938386228\\
1.42857142857143	0.110485519870562\\
1.57142857142857	0.107246615860701\\
1.71428571428571	0.101358298077004\\
1.85714285714286	0.094410220482399\\
2	0.0872514535411281\\
2.14285714285714	0.0803185284460129\\
2.28571428571429	0.0738197415897613\\
2.42857142857143	0.0678383569800374\\
2.57142857142857	0.0623904421555523\\
2.71428571428571	0.0574572990402141\\
2.85714285714286	0.0530035985446782\\
3	0.0489874260330385\\
3.14285714285714	0.0453657311679928\\
3.28571428571429	0.0420971643194595\\
3.42857142857143	0.0391434326150134\\
3.57142857142857	0.0364698270729367\\
3.71428571428571	0.034045296512829\\
3.85714285714286	0.0318422847643186\\
4	0.029836455204528\\
4.14285714285714	0.0280063726800398\\
4.28571428571429	0.0263331813283445\\
4.42857142857143	0.0248002984313796\\
4.57142857142857	0.0233931338083026\\
4.71428571428571	0.0220988382137671\\
4.85714285714286	0.0209060808757388\\
5	0.0198048545675676\\
5.14285714285714	0.0187863057970114\\
5.28571428571429	0.0178425874125187\\
5.42857142857143	0.0169667309372511\\
5.57142857142857	0.0161525361037253\\
5.71428571428571	0.0153944752943099\\
5.85714285714286	0.0146876108481638\\
6	0.0140275234473798\\
6.14285714285714	0.0134102500308495\\
6.28571428571429	0.0128322298974176\\
6.42857142857143	0.0122902578485364\\
6.57142857142857	0.0117814433853187\\
6.71428571428571	0.0113031751172695\\
6.85714285714286	0.0108530896623543\\
7	0.0104290444226995\\
7.14285714285714	0.0100290937095453\\
};
\label{p3}
\addplot [color=mycolor3,dashed,line width=2.0pt,forget plot]
  table[row sep=crcr]{%
-7.14285714285714	0.999950355982668\\
-7	0.999946316562657\\
-6.85714285714286	0.999941861228967\\
-6.71428571428571	0.999936937988365\\
-6.57142857142857	0.999931487180751\\
-6.42857142857143	0.999925440179868\\
-6.28571428571429	0.999918717829202\\
-6.14285714285714	0.999911228594275\\
-6	0.999902866312503\\
-5.85714285714286	0.999893507520433\\
-5.71428571428571	0.999883008196554\\
-5.57142857142857	0.999871199815167\\
-5.42857142857143	0.999857884602701\\
-5.28571428571429	0.999842829608162\\
-5.14285714285714	0.999825759702589\\
-5	0.999806348602295\\
-4.85714285714286	0.999784208270461\\
-4.71428571428571	0.999758875381943\\
-4.57142857142857	0.999729794742806\\
-4.42857142857143	0.999696299030318\\
-4.28571428571429	0.999657582251272\\
-4.14285714285714	0.999612669253118\\
-4	0.999560373424598\\
-3.85714285714286	0.999499249453937\\
-3.71428571428571	0.999427528390758\\
-3.57142857142857	0.999343041580769\\
-3.42857142857143	0.999243127300996\\
-3.28571428571429	0.999124508179037\\
-3.14285714285714	0.998983169345396\\
-3	0.99881418028746\\
-2.85714285714286	0.998611566502406\\
-2.71428571428571	0.998368137961618\\
-2.57142857142857	0.998075477306587\\
-2.42857142857143	0.997724121684024\\
-2.28571428571429	0.997304241165274\\
-2.14285714285714	0.996807515678448\\
-2	0.996230852446075\\
-1.85714285714286	0.995584602266705\\
-1.71428571428571	0.994907457461065\\
-1.57142857142857	0.994294238305226\\
-1.42857142857143	0.993938859599996\\
-1.28571428571429	0.994183147492463\\
-1.14285714285714	0.995491061592862\\
-1	0.998032303207463\\
-0.857142857142857	0.999961934731952\\
-0.714285714285714	0.993156466828522\\
-0.571428571428571	0.963207643880355\\
-0.428571428571429	0.944030108363215\\
-0.285714285714286	0.99462625019553\\
-0.142857142857143	0.742618828360667\\
0	0.398925341067966\\
0.142857142857143	0.742615887394881\\
0.285714285714286	0.994626178572855\\
0.428571428571429	0.94402928919587\\
0.571428571428571	0.9632081694682\\
0.714285714285714	0.993156777941628\\
0.857142857142857	0.999961884746411\\
1	0.998032113363308\\
1.14285714285714	0.995490871868253\\
1.28571428571429	0.994183005993989\\
1.42857142857143	0.993938770499041\\
1.57142857142857	0.994294191311308\\
1.71428571428571	0.99490744059622\\
1.85714285714286	0.995584605283821\\
2	0.996230868031389\\
2.14285714285714	0.99680753858927\\
2.28571428571429	0.997304268006382\\
2.42857142857143	0.997724150194259\\
2.57142857142857	0.998075506054163\\
2.71428571428571	0.99836816608849\\
2.85714285714286	0.998611593468488\\
3	0.998814205834205\\
3.14285714285714	0.998983193328033\\
3.28571428571429	0.999124530576155\\
3.42857142857143	0.999243148135537\\
3.57142857142857	0.999343060916899\\
3.71428571428571	0.999427546312888\\
3.85714285714286	0.999499266051743\\
4	0.999560388795779\\
4.14285714285714	0.999612683489312\\
4.28571428571429	0.999657595444016\\
4.42857142857143	0.999696311263848\\
4.57142857142857	0.999729806096631\\
4.71428571428571	0.999758885929415\\
4.85714285714286	0.999784218078617\\
5	0.999806357733082\\
5.14285714285714	0.999825768211977\\
5.28571428571429	0.999842837547577\\
5.42857142857143	0.999857892018661\\
5.57142857142857	0.999871206750107\\
5.71428571428571	0.99988301468899\\
5.85714285714286	0.999893513605408\\
6	0.999902872021912\\
6.14285714285714	0.999911233957093\\
6.28571428571429	0.999918722871856\\
6.42857142857143	0.999925444926401\\
6.57142857142857	0.999931491653129\\
6.71428571428571	0.999936942206625\\
6.85714285714286	0.999941865211428\\
7	0.999946320326075\\
7.14285714285714	0.999950359542378\\
};
\label{p4}
\end{axis}

\node  at (rel axis cs: 0.75,0.35) {\shortstack[l]{
\ref{p2} $|b^{\rm num.}_{ \scriptscriptstyle T}(\omega)|$ \\
\ref{p3} $|b^{\rm anal.}_{ \scriptscriptstyle T}(\omega)|$}};

\node at (rel axis cs: 0.25,0.7) {\shortstack[l]{
\ref{p1} $|a^{\rm num.}_{ \scriptscriptstyle T}(\omega)|$\\
\ref{p4} $|a^{\rm anal.}_{ \scriptscriptstyle T}(\omega)|$}};
\end{scope}
\end{tikzpicture}

\vspace*{-1.5mm}
\caption{\label{fig:pulse} $(a)$ The truncated 4-soltion pulse $q_{ \scriptscriptstyle T}(t)$, truncated outside $t\in[-3.5,3.5]$ ($T=3.5$). The truncated tails are shown by dashed lines. The tails are approximated by
$\tilde{q}_{ \scriptscriptstyle L}(t)$ and $\tilde{q}_{ \scriptscriptstyle R}(t)$, shown by red dashed. 
The nonlinear spectrum of $q_{ \scriptscriptstyle T}(t)$ is computed and compared numerically and analytically.   
$(b)$ The discrete spectrum (some points are invisible as they overlap) $(c)$ the continuous spectrum.}
\end{figure*}

\begin{theorem}(Continuous Spectrum) When $T$ is large enough, the Jost coefficients 
$(a_{ \scriptscriptstyle T}(\omega),b_{ \scriptscriptstyle T}(\omega))$ are well-approximated 
for any $\omega\in\mathbb{R}$ by
\begin{align}
\label{eq:at_con}
a_{ \scriptscriptstyle T}(\omega) &= a(\omega)(\alpha^*(\omega))^2  - a^*(\omega) (\beta(\omega))^2 e^{-2j\phi} ,\\
b_{ \scriptscriptstyle T}(\omega) &= -\left(a^*(\omega)\alpha(\omega)\beta(\omega)+a(\omega)\alpha^*(\omega)\beta^*(\omega)e^{2j\phi}\right)
\label{eq:bt_con}
\end{align}
where $a(\lambda)$ and $\exp(j\phi)$ are defined in \eqref{eq:a_soliton} and \eqref{eq:b_trunc}.
The continuous spectrum is accordingly obtained by $\tilde{Q}_c(\omega)=\frac{b_{ \scriptscriptstyle T}(\omega)}{a_{ \scriptscriptstyle T}(\omega)}$. 
\label{thm:cont}
\end{theorem}

Now we extend Theorem~\ref{thm:cont} to $\lambda\in\mathbb{C}$  for which the 
Jost coefficients are analytic. 

\begin{theorem}(Nonlinear Fourier Spectrum) Consider a strip of $\lambda\in\mathbb{C}$ 
such that $|\text{Im}(\lambda)|<\sigma_1$. Then, the Jost coefficients 
$(a_{ \scriptscriptstyle T}(\lambda),b_{ \scriptscriptstyle T}(\lambda))$ are well-approximated for large enough $T$ by
\begin{align}
a_{ \scriptscriptstyle T}(\lambda) &= a(\lambda)(\alpha^*(\lambda^*))^2  - a^*(\lambda^*) (\beta(\lambda))^2 e^{-2j\phi}\nonumber\\
 &\hspace*{1.3cm}+ \alpha^*(\lambda^*)\beta(\lambda)\left(b^*(\lambda^*)+b(\lambda)e^{-2j\phi}\right), \label{eq:a_lam}\\
b_{ \scriptscriptstyle T}(\lambda) &= 
-\left(a^*(\lambda^*)\alpha(\lambda)\beta(\lambda)+a(\lambda)\alpha^*(\lambda^*)\beta^*(\lambda^*)e^{2j\phi}\right)\nonumber\\
 &\hspace*{1.3cm} +
\alpha(\lambda)\alpha^*(\lambda^*) b(\lambda) - \beta(\lambda)\beta^*(\lambda^*)b^*(\lambda^*)
e^{2j\phi}\label{eq:b_lam}
\end{align}  
\label{thm:jost_coeff}
\end{theorem}

Note that $a_{ \scriptscriptstyle T}(\lambda)$  and $b_{ \scriptscriptstyle T}(\lambda)$ are entire functions as 
$q_{ \scriptscriptstyle T}(t)$ has a bounded support~\cite{ablowitz1981solitons}. The above equations are the result of \textit{layer-peeling} algorithm and are valid for any $\lambda$. 
When $|\text{Im}(\lambda)|>\sigma_1$, some terms in \eqref{eq:a_lam} and \eqref{eq:b_lam}
become unbounded but they cancel out each other as $a_{ \scriptscriptstyle T}(\lambda)$  and $b_{ \scriptscriptstyle T}(\lambda)$ 
are analytic.

In principle, $(a_{ \scriptscriptstyle T}(\lambda),b_{ \scriptscriptstyle T}(\lambda))$ characterizes 
the entire nonlinear spectrum. 
The continuous spectrum $\tilde{Q}_c(\omega)$ is already given in Theorem~\ref{thm:cont}.  
We discuss next how to find the discrete spectrum. We present first two different methods to locate the eigenvalues, the zeros of $a_{ \scriptscriptstyle T}(\lambda)$. 
Then, we explain how to estimate $b_{\scriptscriptstyle T}(\lambda)$ of each eigenvalue.   
 
\subsection{Finding Discrete Eigenvalues from Continuous Spectrum}\label{sec:ev_con}

It is usually much easier to numerically compute the continuous spectrum than the discrete spectrum. Equation \eqref{eq:ZS} is skew-Hermitian for $\lambda\in\mathbb{R}$, while it becomes even unbounded for $\lambda$ with a large imaginary part. Here,
we present how to find the discrete eigenvalues from the continuous spectrum.

Consider an arbitrary pulse with the known Jost coefficients $(a(\omega), b(\omega))$ for all $\omega\in\mathbb{R}$.
The relation between $a(\lambda)$ and $b(\omega))$ is given in \eqref{eq:a_from_b}. 
When $\lambda=\omega\in\mathbb{R}$, this relation  changes slightly to  \cite[P. 49, Eq. 6.27]{faddeev2007hamiltonian} 
\begin{equation*}
a(\omega) = \mathsf{a}[b(\omega)]\prod_{k=1}^N\frac{\omega-\lambda_k}{\omega-\lambda_k^*}
\end{equation*}
where the functional
\begin{equation*}
\mathsf{a}[b(\omega)]=\sqrt{1-|b(\omega)|^2}\exp\left(\frac{j}{2} \mathcal{H}\left[\ln(1-|b(\omega)|^2)\right]\right)
\end{equation*}
and $\mathcal{H}[\cdot]$ denotes the Hilbert transform. 
Note that $\mathsf{a}[b(\omega)]$ is the $a$-coefficient of the non-solitonic part of the pulse. Define,
\begin{equation}\label{eq:allpass}
G(\omega)=\frac{a(\omega)}{\mathsf{a}[b(\omega)]} = \prod_{k=1}^N\frac{\omega-\lambda_k}{\omega-\lambda_k^*}.
\end{equation}
Knowing $(a(\omega),b(\omega))$, we can find $G(\omega)$ which is an all-pass filter, $|G(\omega)|=1$ and its phase diagram provides the location of eigenvalues. The first result is on the number of eigenvalues:
\begin{corollary}
Let $\theta(\omega)=-j\ln(G(\omega))$. The number of eigenvalues, $N$, is equal to the number of times $e^{j\theta(\omega)}$
encircles the origin when $\omega$ traverses from $-\infty$ to $\infty$. In other words, $\lim_{\omega\to\infty} \theta(\omega)-\theta(-\omega)=2N\pi$.
\end{corollary}

Theorem~\ref{thm:cont} gives $(a_{ \scriptscriptstyle T}(\omega),b_{ \scriptscriptstyle T}(\omega))$. Accordingly,
we estimate $\mathsf{a}[b_{ \scriptscriptstyle T}(\omega)]$ and $G(\omega)$.
Knowing $N$, the number of eigenvalues, $G(\omega)$ has $2N$ unknown variables 
which are detectable from at least $2N$ distinct frequencies of $G(\omega)$. 
However, $(a_{ \scriptscriptstyle T}(\omega),b_{ \scriptscriptstyle T}(\omega))$ should be known in much more frequencies not only to 
compute $\mathsf{a}[b_{ \scriptscriptstyle T}(\omega)]$ but also 
to overcome the algorithmic numerical errors of $(a_{ \scriptscriptstyle T}(\omega),b_{ \scriptscriptstyle T}(\omega))$. Phase synthesis of an $N-$order all pass filter is a classical signal-processing problem, e.g. \cite{lang1994simple} and references therein. 
Here, we used a sub-optimal least mean squared error based estimation method. It gives almost the same result
as the next method.

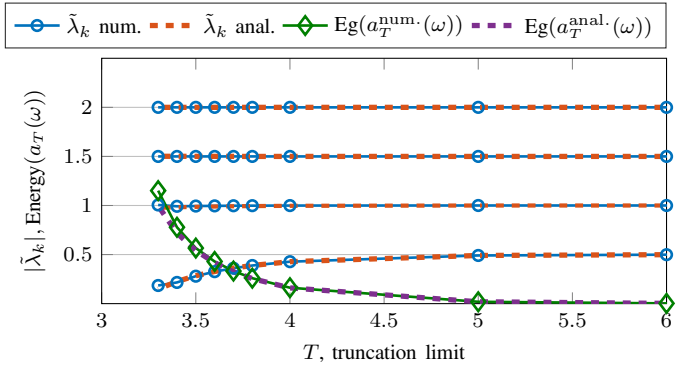
\begin{figure}[t!]
\setlength{\wlength}{0.43\textwidth}
\setlength{\hlength}{0.18\textwidth}
\begin{center}

\definecolor{mycolor1}{rgb}{0.00000,0.44706,0.74118}%
\definecolor{mycolor2}{rgb}{0.85098,0.32549,0.09804}%
\definecolor{mycolor3}{rgb}{0.00000,0.49804,0.00000}%
\definecolor{mycolor4}{rgb}{0.49412,0.18431,0.55686}%
\begin{tikzpicture}[every node/.style={font=\footnotesize}]

\begin{axis}[%
width=0.963\wlength,
height=\hlength,
at={(0\wlength,0\hlength)},
scale only axis,
xmin=3,
xmax=6,
xlabel={$T$, truncation limit},
ymin=0,
ymax=2.5,
ytick={  0.5, 1,    1.5,  2},
ylabel={$|\tilde{\lambda}_k|, \text{Energy}(a_{ \scriptscriptstyle T}(\omega))$},
ymajorgrids={true},
axis background/.style={fill=white},
legend style={legend cell align=left,align=left,draw=white!15!black,at={(axis cs:6,2.6)},anchor=south east,legend columns=4,font=\footnotesize}
]
\addplot [color=mycolor1,solid,line width=1.0pt,mark=o,mark options={solid,fill=mycolor1},forget plot]
  table[row sep=crcr]{%
8	0.499977105279275\\
7	0.499833302065798\\
6	0.498766768135025\\
5	0.490760519630527\\
4	0.427359639150621\\
3.8	0.388400926128664\\
3.7	0.36094816104802\\
3.6	0.325750694487693\\
3.5	0.279587262269632\\
3.4	0.215958051316197\\
3.3	0.184815751656623\\
};
\addplot [color=mycolor2,dashed,line width=2pt,forget plot]
  table[row sep=crcr]{%
8	0.499977762058661\\
7	0.499839522979435\\
6	0.498723331378039\\
5	0.490769757181957\\
4	0.427454832695925\\
3.8	0.3887864109588\\
3.7	0.361869994792526\\
3.6	0.3275\\
3.5	0.284002955453648\\
3.4	0.225928857443635\\
3.3	0.1475\\
};
\addplot [color=mycolor1,solid,line width=1.0pt,mark=o,mark options={solid,fill=mycolor1},forget plot]
  table[row sep=crcr]{%
8	1.00000037918792\\
7	1.00000029542143\\
6	1.00000012879644\\
5	0.999987759305029\\
4	0.999184408896717\\
3.8	0.998194567510027\\
3.7	0.997334723226782\\
3.6	0.99608998449469\\
3.5	0.994304991433164\\
3.4	0.991772343280159\\
3.3	1.00537813144786\\
};
\addplot [color=mycolor2,dashed,line width=1.6pt,forget plot]
  table[row sep=crcr]{%
8	0.999999972316615\\
7	0.999999444084247\\
6	0.999991328119767\\
5	0.999819102890233\\
4	0.99658519958841\\
3.8	0.993973016113619\\
3.7	0.992026601466366\\
3.6	0.989486585025898\\
3.5	0.98618939389479\\
3.4	0.981933527854742\\
3.3	0.976476096335175\\
};
\addplot [color=mycolor1,solid,line width=1.0pt,mark=o,mark options={solid,fill=mycolor1},forget plot]
  table[row sep=crcr]{%
8	1.50000180802327\\
7	1.50000138417836\\
6	1.500001016686\\
5	1.50000069773188\\
4	1.49999966089632\\
3.8	1.49999770464699\\
3.7	1.49999529067514\\
3.6	1.49999071034429\\
3.5	1.49998204812359\\
3.4	1.49996578965354\\
3.3	1.49992813219936\\
};
\addplot [color=mycolor2,dashed,line width=2.0pt,forget plot]
  table[row sep=crcr]{%
8	1.49999999998201\\
7	1.49999999901791\\
6	1.4999999464367\\
5	1.50000202485236\\
4	1.50011093514479\\
3.8	1.49976047651259\\
3.7	1.50034820064813\\
3.6	1.50050738992792\\
3.5	1.50073469846573\\
3.4	1.5010577281696\\
3.3	1.50151265816351\\
};
\addplot [color=mycolor1,solid,line width=1.0pt,mark=o,mark options={solid,fill=mycolor1}]
  table[row sep=crcr]{%
8	2.00001235041241\\
7	2.00000945587722\\
6	2.00000694723096\\
5	2.00000482423868\\
4	2.00000304950978\\
3.8	2.00000268419014\\
3.7	2.00000247449101\\
3.6	2.00000222799019\\
3.5	2.00000191964877\\
3.4	2.00000150928309\\
3.3	2.00000049769907\\
};
\addlegendentry{$\tilde{\lambda}_k$ num.};
\addplot [color=mycolor2,dashed,line width=2.0pt]
  table[row sep=crcr]{%
8	2\\
7	2.00000000000053\\
6	2.00000000007852\\
5	1.99999998842174\\
4	2.00000109521952\\
3.8	2.00000287621196\\
3.7	2.00000442462574\\
3.6	2.00000753280608\\
3.5	1.99998719360236\\
3.4	2.00001962923113\\
3.3	1.99996823637222\\
};
\addlegendentry{$\tilde{\lambda}_k$ anal.};
\addplot [color=mycolor3,solid,line width=1.0pt,mark size=3.5pt,mark=diamond,mark options={solid,fill=mycolor3}]
  table[row sep=crcr]{%
8	4.42942149897629e-05\\
7	0.000329278209709435\\
6	0.00245950397703397\\
5	0.0188833120396623\\
4	0.162131017220375\\
3.8	0.25924451801644\\
3.7	0.33180429318756\\
3.6	0.429479366974058\\
3.5	0.566111271390382\\
3.4	0.778276829838544\\
3.3	1.1511963752935\\
};
\addlegendentry{Eg$(a^{\rm num.}_{ \scriptscriptstyle T}(\omega))$};
\addplot [color=mycolor4,dashed,line width=2.0pt]
  table[row sep=crcr]{%
8	4.42961062203372e-05\\
7	0.000329311533694126\\
6	0.00245980597410863\\
5	0.018867654555985\\
4	0.160430947686393\\
3.8	0.254866091161142\\
3.7	0.324301756971203\\
3.6	0.416368534711691\\
3.5	0.541377024442223\\
3.4	0.7184928651108\\
3.3	0.996314839555906\\
};
\addlegendentry{Eg$(a^{\rm anal.}_{ \scriptscriptstyle T}(\omega))$};
\end{axis}
\end{tikzpicture}%
\caption{\label{fig:EV_average} The effect of truncation of 4-soliton pulses on their eigenvalues and the continuous spectrum. We compare the analytic estimates of the eigenvalues with the numerically averaged eigenvalues as well as the energy in continuous spectrum from the analytic estimation with the averaged energy from the numerical computations. All the curves overlap almost at every point.}
\end{center}
\end{figure}

\subsection{Zero-searching of Eigenvalues}\label{sec:zero_search}

Let $\tilde{\lambda}_k$ denote the eigenvalues of $a_{ \scriptscriptstyle T}(\lambda)$, i.e. $a_{ \scriptscriptstyle T}(\tilde{\lambda}_k)=0$. Assume that $\text{Im}(\tilde{\lambda}_k)\leq \text{Im}(\tilde{\lambda}_{k+1})$ for all $k$. Define
\begin{equation}\label{eq:a_lam2}
\tilde{a}_{ \scriptscriptstyle T}(\lambda)= a(\lambda)(\alpha^*(\lambda^*))^2  - a^*(\lambda^*) (\beta(\lambda))^2 e^{-2j\phi}
\end{equation} 
and $a^*(\lambda^*)=1/a(\lambda)$ given in \eqref{eq:a_soliton}. The above function is analytic
for $\lambda\in\mathbb{C}^+\backslash\{\lambda_k\}_{k=1}^N$. 
From \eqref{eq:b_sol}, $\tilde{a}_{ \scriptscriptstyle T}(\lambda)=a_{ \scriptscriptstyle T}(\lambda)$  when $0\leq\text{Im}(\lambda)<\sigma_1$. Therefore, we estimate the eigenvalues by zeros of $\tilde{a}_{ \scriptscriptstyle T}(\lambda)$
which satisfy
\begin{equation}\label{eq:zero_search}
a(\lambda)^2-\left(\frac{\beta(\lambda)e^{-j\phi}}{\alpha^*(\lambda^*)}\right)^2=0.
\end{equation}

Using \eqref{eq:a_soliton}, \eqref{eq:a_trunc} and \eqref{eq:b_trunc}, we find the zeros of the above equation (a numerical zero-search) in the next section.
We will observe that the estimated eigenvalues are rather precise. 

Estimation of $b_{ \scriptscriptstyle T}(\tilde{\lambda}_k)$ is more challenging. If $\text{Im}(\tilde{\lambda}_k)<\sigma_1$ (it is the case for $\tilde{\lambda}_1$), $b(\tilde{\lambda}_k)=b^*(\tilde{\lambda}^*_k)=0$ and thus,
\begin{equation*}
b_{ \scriptscriptstyle T}(\tilde{\lambda}_k) \approx-\left(a(\tilde{\lambda}_k)^{-1}\alpha(\tilde{\lambda}_k)\beta(\tilde{\lambda}_k)+a(\tilde{\lambda}_k)\alpha^*(\tilde{\lambda}^*_k)\beta^*(\tilde{\lambda}^*_k)e^{2j\phi}\right).
\end{equation*}

When $\text{Im}(\tilde{\lambda}_k)>\sigma_1$,
 $b(\tilde{\lambda}_k)$ and $b^*(\tilde{\lambda}^*_k)$ are unknown and can be even unbounded.
Assuming that the eigenvalues after truncation are slightly perturbed ($\tilde{\lambda}_k\approx \lambda_k$), we set
$b(\tilde{\lambda}_k)\approx b(\lambda_k)$ and $b^*(\tilde{\lambda}^*_k)\approx b^*(\lambda^*_k)$. Note that for the original multi-soliton, $b^*(\lambda^*_k)=1/b(\lambda_k)$ at its eigenvalue $\lambda_k$. In this case,
 \begin{equation*}
b_{ \scriptscriptstyle T}(\tilde{\lambda}_k) \approx
\alpha(\tilde{\lambda}_k)\alpha^*(\tilde{\lambda}^*_k) b(\lambda_k) - \beta(\tilde{\lambda}_k)\beta^*(\tilde{\lambda}^*_k)
e^{2j\phi}\frac{1}{b(\lambda_k)},
\end{equation*}
In the next section, we will show numerically that $\{\tilde{\lambda}_k,b_{ \scriptscriptstyle T}(\tilde{\lambda}_k)\}$ are estimated rather precise.

\section{Numerical Evaluation}\label{sec:simul}

We verified the precision of our analytic estimations numerically. Consider 
multi-soliton pulses with 4 eigenvalues $\{\lambda_1=0.5j,\lambda_2=1j,\lambda_3=1.5j,\lambda_4=2j\}$.
Letting  $b(\lambda_k)=\exp(j\phi_k)$, we chose each $\phi_k$ randomly. We generated 1000 such pulses for our
comparison. Each pulse was truncated in a window of $|t|\leq T$ for different values of $T$. 
After truncation, we computed the nonlinear spectrum numerically. 
To have small numerical errors, each pulse was uniformly sampled by 10000 points. The continuous spectrum, denoted by
$(a^{\text{num.}}_{ \scriptscriptstyle T}(\omega),b^{\text{num.}}_{ \scriptscriptstyle T}(\omega))$, is computed numerically using the forward-backward algorithm~\cite{aref2016control} (forward part is enough), 
which gave more precise results than Boffetta-Osborne algorithm~\cite{boffetta1992computation}. 
The discrete part, denoted by
$\{\tilde{\lambda}^{\rm num.}_k,b^{\rm num.}_{ \scriptscriptstyle T}(\tilde{\lambda}_k)\}$ was computed 
using the forward-backward algorithm~\cite{aref2016control}: 
The eigenvalues $\tilde{\lambda}^{\rm num.}_k$ were found by Newton-Raphson 
zero-search method with stopping condition $|a_{ \scriptscriptstyle T}(\tilde{\lambda}^{\rm num.}_k)|<10^{-6}$. 

The analytic estimates were computed according to Theorem~\ref{thm:cont} for the continuous spectrum,
denoted here by $(a^{\text{anal.}}_{ \scriptscriptstyle T}(\omega),b^{\text{anal.}}_{ \scriptscriptstyle T}(\omega))$, and according to Section~\ref{sec:zero_search} for the discrete spectrum, denoted here by  
$\{\tilde{\lambda}^{\rm anal.}_k,b^{\rm anal.}_{ \scriptscriptstyle T}(\tilde{\lambda}_k)\}$.

Fig.~\ref{fig:pulse} shows an example of such pulses with its spectrum after truncation. The pulse
is truncated in the window of $|t|\leq 3.5$. The truncated tails are shown in dashed lines. We observe that
the analytic estimates of the nonlinear spectrum closely follow its numerical estimates.
To measure the estimation precision, we compute the estimation errors averaged over 1000 randomly generated pulses.
Fig.~\ref{fig:EV_average} illustrates the mean value of $\tilde{\lambda}^{\rm num.}_k$ in terms of $T$, truncation
interval. We also plot the energy of the continuous spectrum, obtained by~\cite{ablowitz1981solitons}
\begin{equation*}
\text{Eg}[a_{ \scriptscriptstyle T}(\omega)]=-\frac{1}{\pi}\int \ln\left(|a_{ \scriptscriptstyle T}(\omega)|^2\right)\text{d}\omega
\end{equation*}

Fig.~\ref{fig:EV_average} shows how much the truncation perturbs the designed eigenvalues, specially $\lambda_1$, and shows the exponential growth of energy in the continuous spectrum.
We observe that the analytic estimates match precisely with the averaged numerical estimates. 
Note that for these pulses, an aggressive truncation with $T<3.2$ may cause missing of one eigenvalue and a larger variations in the remaining 3 eigenvalues. At this limit, our analytic estimates are not precise anymore.

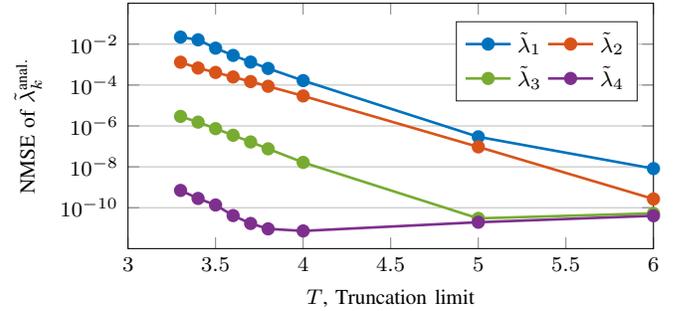
\begin{figure}[t!]
\setlength{\wlength}{0.4\textwidth}
\setlength{\hlength}{0.18\textwidth}
\begin{center}
\definecolor{mycolor1}{rgb}{0.00000,0.44700,0.74100}%
\definecolor{mycolor2}{rgb}{0.00000,0.44706,0.74118}%
\definecolor{mycolor3}{rgb}{0.85000,0.32500,0.09800}%
\definecolor{mycolor4}{rgb}{0.85098,0.32549,0.09804}%
\definecolor{mycolor5}{rgb}{0.46667,0.67451,0.18824}%
\definecolor{mycolor6}{rgb}{0.49400,0.18400,0.55600}%
\definecolor{mycolor7}{rgb}{0.49412,0.18431,0.55686}%
\begin{tikzpicture}[every node/.style={font=\footnotesize}]

\begin{axis}[%
width=0.963\wlength,
height=\hlength,
at={(0\wlength,0\hlength)},
scale only axis,
xmin=3,
xmax=6,
xlabel={$T$, Truncation limit},
ymode=log,
ymin=1e-12,
ymax=1,
yminorticks=true,
ytick={ 1e-2,    1e-4,  1e-6,    1e-8,1e-10},
ymajorgrids={true},
ylabel={NMSE of $\tilde{\lambda}^{\text{anal.}}_k$},
axis background/.style={fill=white},
legend style={at={(0.97,0.60)},anchor=south east,legend columns=2,legend cell align=left,align=left,draw=white!15!black, font=\footnotesize}
]
\addplot [color=mycolor1,solid,line width=1.0pt,mark=*,mark options={solid,fill=mycolor2}]
  table[row sep=crcr]{%
8	6.5720738874168e-12\\
7	1.57636960154761e-10\\
6	8.20187634908755e-09\\
5	2.9314121127241e-07\\
4	0.000165138790531848\\
3.8	0.00064870009487767\\
3.7	0.00132862190199907\\
3.6	0.00281751129965045\\
3.5	0.00637053403249952\\
3.4	0.0163138591586289\\
3.3	0.0221637608036723\\
};
\addlegendentry{$\tilde{\lambda}_1$};

\addplot [color=mycolor3,solid,line width=1.0pt,mark=*,mark options={solid,fill=mycolor4}]
  table[row sep=crcr]{%
8	4.78242547521206e-11\\
7	2.9662591616905e-11\\
6	2.68058605103183e-10\\
5	9.42200342061864e-08\\
4	2.936159379725e-05\\
3.8	8.69783996057661e-05\\
3.7	0.000147777508122577\\
3.6	0.000248602452349246\\
3.5	0.00041370086027171\\
3.4	0.000680386894174832\\
3.3	0.00131052299834305\\
};
\addlegendentry{$\tilde{\lambda}_2$};

\addplot [color=mycolor5,solid,line width=1.0pt,mark=*,mark options={solid,fill=mycolor5}]
  table[row sep=crcr]{%
8	1.67504867501134e-10\\
7	9.81995366435793e-11\\
6	5.30562401047407e-11\\
5	3.0098137471757e-11\\
4	1.64328464450706e-08\\
3.8	7.59834172457074e-08\\
3.7	1.64360614622319e-07\\
3.6	3.49798261337498e-07\\
3.5	7.36954162183793e-07\\
3.4	1.53609937468171e-06\\
3.3	2.92981030244909e-06\\
};
\addlegendentry{$\tilde{\lambda}_3$};

\addplot [color=mycolor6,solid,line width=1.0pt,mark=*,mark options={solid,fill=mycolor7}]
  table[row sep=crcr]{%
8	1.27766983435427e-10\\
7	7.49014368508169e-11\\
6	4.04328311008617e-11\\
5	1.95294238773761e-11\\
4	7.33025048365381e-12\\
3.8	9.2643811875211e-12\\
3.7	1.69830849833168e-11\\
3.6	4.1259300600661e-11\\
3.5	1.35713189459557e-10\\
3.4	2.84251003365671e-10\\
3.3	7.02337694429448e-10\\
};
\addlegendentry{$\tilde{\lambda}_4$};

\end{axis}
\end{tikzpicture}%
\caption{\label{fig:EV_MSE} Normalized mean square error (NMSE) of eigenvalues analytic estimation in terms of truncation limit.}
\end{center}
\end{figure}

\begin{figure}[t!]
\setlength{\wlength}{0.4\textwidth}
\setlength{\hlength}{0.18\textwidth}
\begin{center}
\definecolor{mycolor1}{rgb}{0.00000,0.44706,0.74118}%
\definecolor{mycolor2}{rgb}{0.85000,0.32500,0.09800}%
\definecolor{mycolor3}{rgb}{0.85098,0.32549,0.09804}%
\definecolor{mycolor4}{rgb}{0.46667,0.67451,0.18824}%
\definecolor{mycolor5}{rgb}{0.49400,0.18400,0.55600}%
\definecolor{mycolor6}{rgb}{0.49412,0.18431,0.55686}%
\begin{tikzpicture}[every node/.style={font=\footnotesize}]

\begin{axis}[%
width=0.963\wlength,
height=\hlength,
at={(0\wlength,0\hlength)},
scale only axis,
xmin=3,
xmax=6,
xlabel={$T$, truncation limit},
ymode=log,
ymin=1e-12,
ymax=1,
yminorticks=true,
ytick={ 1e-2,    1e-4,  1e-6,    1e-8,1e-10},
ymajorgrids={true},
ylabel={NMSE of  $b_{ \scriptscriptstyle T}(\tilde{\lambda}_k)$},
axis background/.style={fill=white},
legend style={at={(0.,0.)},anchor=south west,legend columns=1, legend cell align=left,align=left,draw=none, font=\footnotesize}
]
\addplot [color=mycolor1,solid,line width=1.0pt,mark=*,mark options={solid,fill=mycolor1}]
  table[row sep=crcr]{%
8	0.000146355051812769\\
7	0.00131545294915756\\
6	0.0001132976539979\\
5	0.000262478416066717\\
4	0.00488782808220946\\
3.8	0.00782433934936722\\
3.7	0.00956346307485915\\
3.6	0.0116193174190739\\
3.5	0.0132122915959919\\
3.4	0.0145021695246919\\
3.3	0.01519988284354\\
};
\label{p11}

\addplot [color=mycolor2,solid,line width=1.0pt,mark=*,mark options={solid,fill=mycolor3}]
  table[row sep=crcr]{%
8	5.25404845443503e-10\\
7	2.48110503905775e-08\\
6	1.35253712174199e-06\\
5	7.44328665310491e-05\\
4	0.00391134769250564\\
3.8	0.00840007161936506\\
3.7	0.0122182838914642\\
3.6	0.0176620730266365\\
3.5	0.0253483530389159\\
3.4	0.0360829611678086\\
3.3	0.0488557503130308\\
};
\label{p21}

\addplot [color=mycolor4,solid,line width=1.0pt,mark=*,mark options={solid,fill=mycolor4}]
  table[row sep=crcr]{%
8	2.18571776944059e-10\\
7	3.5273685234823e-09\\
6	1.87556957920087e-07\\
5	1.0144424588061e-05\\
4	0.000517068694091404\\
3.8	0.00110941659536341\\
3.7	0.00161213781229391\\
3.6	0.00233413279159511\\
3.5	0.00335981278806709\\
3.4	0.00480347925193219\\
3.3	0.00712032576683979\\
};
\label{p31}

\addplot [color=mycolor5,solid,line width=1.0pt,mark=*,mark options={solid,fill=mycolor6}]
  table[row sep=crcr]{%
8	1.27163192551214e-10\\
7	1.06820795926585e-09\\
6	5.45184077115997e-08\\
5	2.93449510271191e-06\\
4	0.000145557759013865\\
3.8	0.000307234846925904\\
3.7	0.000442629961142342\\
3.6	0.000633257992796145\\
3.5	0.0008986589013198\\
3.4	0.00126314491895767\\
3.3	0.00180459285808461\\
};
\label{p41}

\addplot [color=mycolor1,dashed,line width=1.0pt,mark=*,mark options={solid,fill=mycolor1}]
  table[row sep=crcr]{%
8	3.76117234918542e-12\\
7	4.88733834138823e-11\\
6	1.18353002859212e-08\\
5	2.7018629935371e-06\\
4	0.000431999700010635\\
3.8	0.0010760591270171\\
3.7	0.00165922924141642\\
3.6	0.00250793116455407\\
3.5	0.00369576405891379\\
3.4	0.00525603640153202\\
3.3	0.00869471404253679\\
};
\label{p51}

\end{axis}
\node[font=\footnotesize]  at (rel axis cs: 0.43,1.25) {\shortstack[l]{
\ref{p51} NMSE of arg$(b_{ \scriptscriptstyle T}(\tilde{\lambda}_1))$\\
\ref{p11} NMSE of $b_{ \scriptscriptstyle T}(\tilde{\lambda}_1)$ \ref{p21} NMSE of $b_{ \scriptscriptstyle T}(\tilde{\lambda}_2)$ \\
\ref{p31} NMSE of $b_{ \scriptscriptstyle T}(\tilde{\lambda}_3)$ \ref{p41} NMSE of $b_{ \scriptscriptstyle T}(\tilde{\lambda}_4)$ }};

\end{tikzpicture}%
\caption{\label{fig:b_MSE} Normalized mean square error (NMSE) of $b^{\rm anal.}_{ \scriptscriptstyle T}(\tilde{\lambda}_k)$ in terms of truncation limit.}
\end{center}
\end{figure}
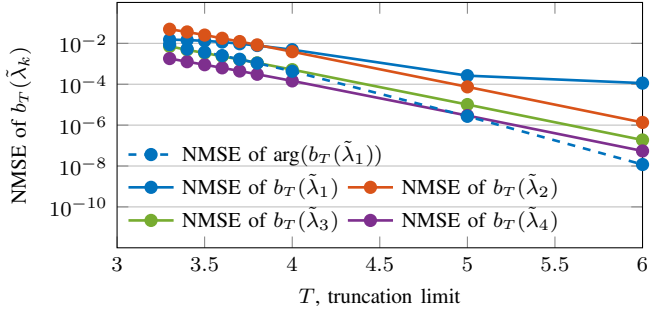

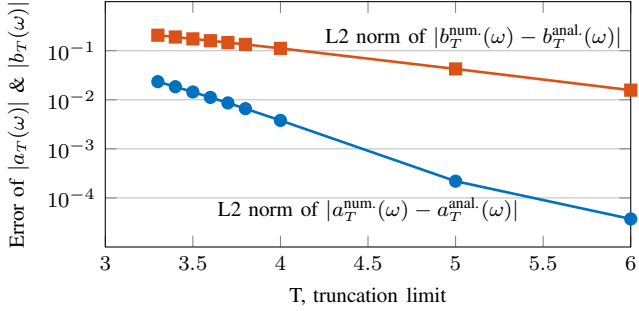
\begin{figure}[t!]
\setlength{\wlength}{0.4\textwidth}
\setlength{\hlength}{0.18\textwidth}
\begin{center}
\definecolor{mycolor1}{rgb}{0.00000,0.44700,0.74100}%
\definecolor{mycolor2}{rgb}{0.00000,0.44706,0.74118}%
\definecolor{mycolor3}{rgb}{0.85000,0.32500,0.09800}%
\definecolor{mycolor4}{rgb}{0.85098,0.32549,0.09804}%
\begin{tikzpicture}[every node/.style={font=\footnotesize}]

\begin{axis}[%
width=0.963\wlength,
height=\hlength,
at={(0\wlength,0\hlength)},
scale only axis,
xmin=3,
xmax=6,
xlabel={T, truncation limit},
ymode=log,
ymin=1e-5,
ymax=1,
yminorticks=true,
ytick={ 1e-1,    1e-2,  1e-3,    1e-4},
ymajorgrids={true},
ylabel={Error of $|a_{ \scriptscriptstyle T}(\omega)|$ \& $|b_{ \scriptscriptstyle T}(\omega)|$},
axis background/.style={fill=white},
legend style={at={(0.505,0.764)},anchor=south west,legend cell align=left,align=left,draw=white!15!black}
]
\addplot [color=mycolor1,solid,line width=1.0pt,mark=*,mark options={solid,fill=mycolor2}]
  table[row sep=crcr]{%
8  0.000045400000293\\
7   0.000040356952543\\
6   0.000037315252301\\
5   0.000219357486257\\
4   0.003800103848905\\
3.8   0.006595459330194\\
3.7   0.008625230303476\\
3.6   0.011179943407300\\
3.5   0.014435539430763\\
3.4   0.018452556211744\\
3.3   0.023429067993896\\
};

\addplot [color=mycolor3,solid,line width=1.0pt,mark=square*,mark options={solid,fill=mycolor4}]
  table[row sep=crcr]{%
8.0  0.002289711601055\\
7.0   0.005780380389308\\
6.0   0.015712497194162\\
5.0   0.042518865213822\\
4.0   0.111728823750487\\
3.8    0.134371367095640\\
3.7    0.146486451248822\\
3.6    0.159873552433434\\
3.5    0.174306262794395\\
3.4    0.190212490760706\\
3.3    0.207150793232696\\
};
\end{axis}
\node at (4.9,2.8) {L2 norm of $|b^{\text{num.}}_{ \scriptscriptstyle T}(\omega)-b^{\text{anal.}}_{ \scriptscriptstyle T}(\omega)|$};
\node at (3.5,0.5) {L2 norm of $|a^{\text{num.}}_{ \scriptscriptstyle T}(\omega)-a^{\text{anal.}}_{ \scriptscriptstyle T}(\omega)|$};

\end{tikzpicture}%
\caption{\label{fig:con_l2} L2 norm of error between analytic estimation and numerical estimation of continuous spectrum in terms of truncation limit.}
\end{center}
\end{figure}

Although the mean numerical estimates are very close to the analytic estimates, the nonlinear spectrum of each pulse fluctuates slightly around its mean values. We compute the normalized mean square error (NMSE) of discrete spectrum in Fig.~\ref{fig:EV_MSE} and in Fig.~\ref{fig:b_MSE}. For the eigenvalues, the NMSE is defined as 
$\frac{1}{|\lambda_k^2|}\mathbb{E}(|\tilde{\lambda}^{\text{anal.}}_k-\tilde{\lambda}^{\text{num.}}_k|^2)$ and for the norming coefficeint, it is defined as $\mathbb{E}(|\frac{b^{\rm anal.}_{ \scriptscriptstyle T}(\tilde{\lambda}_k)}{b^{\rm num.}_{ \scriptscriptstyle T}(\tilde{\lambda}_k)}-1|^2)$. 

Fig.~\ref{fig:con_l2} shows the mean accumulated error in estimation of the continuous spectrum. The error is computed by the L2-norms $\parallel b^{\text{num.}}_{ \scriptscriptstyle T}(\omega)-b^{\text{anal.}}_{ \scriptscriptstyle T}(\omega)\parallel_2$ and $\parallel a^{\text{num.}}_{ \scriptscriptstyle T}(\omega)-a^{\text{anal.}}_{ \scriptscriptstyle T}(\omega)\parallel_2$ over a wide range of $\omega$, e.g. see Fig.~\ref{fig:pulse}$(c)$.
Note that the average error per sample $\omega$ is an order of magnitude smaller.

\section{Sketch of the Proof of Theorem \ref{thm:jost_coeff}}\label{sec:anal}
Theorem~\ref{thm:cont} is a special case of Theorem~\ref{thm:jost_coeff} and therefore, we 
prove the latter.
The proof is rather general and can be applied for any pulse and different truncation boundaries. 
It results, however, simple relations for symmetric multi-solitons.

Consider a multi-soliton pulse $q(t)$ with $\{\lambda_k,b(\lambda_k)\}_{k=1}^N$ such that $\lambda_1=\omega_1+j\sigma_1$
and $\sigma_1<\text{Im}(\lambda_k)$. Without loss of generality, we assume that $|b(\lambda_1)|=1$. 
This condition is equivalent to translate the pulse by a constant time shift.
The pulse $q(t)$ can be decomposed into three parts with disjoint supports (see Fig.~\ref{fig:pulse}): 
its left tail denoted by $q_{ \scriptscriptstyle L}(t)$, 
middle part $q_{ \scriptscriptstyle T}(t)$ and 
the right tail, denoted by $q_{ \scriptscriptstyle R}(t)$.
The proof has the following steps:
\begin{itemize}
\item[1)] Derive the Jost pair $(a_{ \scriptscriptstyle L}(\lambda),b_{ \scriptscriptstyle L}(\lambda))$ 
of the left tail $q_{ \scriptscriptstyle L}(t)$.
\item[2)] Derive the Jost pair $(a_{ \scriptscriptstyle R}(\lambda),b_{ \scriptscriptstyle R}(\lambda))$
of the right tail $q_{ \scriptscriptstyle R}(t)$.
\item[3)] Using \textit{layer-peeling} method~\cite{yousefi2014information} to express  
$(a_{ \scriptscriptstyle T}(\lambda),b_{ \scriptscriptstyle T}(\lambda))$ 
in terms of Jost pairs of $q(t)$, $q_{ \scriptscriptstyle L}(t)$ and $q_{ \scriptscriptstyle R}(t)$.
\end{itemize}

We showed in \cite{span2017timebandwidth} that $q_{ \scriptscriptstyle L}(t)$ and $q_{ \scriptscriptstyle R}(t)$
can be approximated
\begin{align*}
\tilde{q}_{ \scriptscriptstyle L}(t)
& =
-2\sigma_1 e^{-j\phi_L-2j\omega_1t}\text{sech}(2\sigma_1(t-t_L)), \text{when }t<t_L\\
\tilde{q}_{ \scriptscriptstyle R}(t)
&= -2\sigma_1 e^{-j\phi_R-2j\omega_1t}\text{sech}(2\sigma_1(t-t_R)), \text{when }t>t_R
\end{align*}
with the parameters $t_R=-t_L=t_0$ and,
\begin{align*}
\phi_{0} &= \sum_{k=2}^N\arg\left(\frac{\lambda_1-\lambda_k^*}{\lambda_1-\lambda_k}\right), 
t_0=\frac{1}{2\sigma_1}\sum_{k=2}^N\ln\left(\left|\frac{\lambda_1-\lambda_k^*}{\lambda_1-\lambda_k}\right|\right)\\
\phi_L&=\arg(b(\lambda_1)) - \phi_0, \hspace{0.45cm}\phi_R=\arg(b(\lambda_1)) + \phi_0
\end{align*}
The approximations are practically very precise,
 specially when $|t|\gg t_0$.
\subsubsection{ Finding $(a_{ \scriptscriptstyle L}(\lambda),b_{ \scriptscriptstyle L}(\lambda))$} 
Now consider $\tilde{q}_{ \scriptscriptstyle L}(t)$. One can verify that the solution of \eqref{eq:ZS} is 
\begin{align*}
v_1(t)&=1-\frac{e^{2\sigma_1(t-t_L)}}{e^{2\sigma_1(t-t_L)}+e^{-2\sigma_1(t-t_L)}}\left(\frac{2j\sigma_1}{\lambda-\omega_1+j\sigma_1}\right)\\
v_2(t)&=\left(\frac{2j\sigma_1}{\lambda-\omega_1+j\sigma_1}\right)
\frac{e^{j\phi_L+2j\omega_1 t}}{e^{2\sigma_1(t-t_L)}+e^{-2\sigma_1(t-t_L)}}
e^{-2j\lambda t}
\end{align*}
provided that $\text{Im}(\lambda)>-\sigma_1$. Assuming $T>t_0$, 
$q_{ \scriptscriptstyle L}(t)\approx\tilde{q}_{ \scriptscriptstyle L}(t)$ if $t\leq -T$ and zero, otherwise. Therefore, 
\begin{equation}\label{eq:v1v2}
(a_{ \scriptscriptstyle L}(\lambda),b_{ \scriptscriptstyle L}(\lambda))
\approx
(v_1(-T),v_2(-T))
\end{equation}
For symmetric multi-solitons in Lemma~\ref{lem:symmetric_sol}, it simplifies to
\begin{equation}\label{eq:left_scattering}
(a_{ \scriptscriptstyle L}(\lambda),b_{ \scriptscriptstyle L}(\lambda))
\approx
(\alpha(\lambda),\beta(\lambda))
\end{equation}
where $\alpha(\lambda)$ and $\beta(\lambda)$ are defined in \eqref{eq:a_trunc} and \eqref{eq:b_trunc}.
%
\subsubsection{ Finding $(a_{ \scriptscriptstyle R}(\lambda),b_{ \scriptscriptstyle R}(\lambda))$} 
Similar to the case of $\tilde{q}_{ \scriptscriptstyle L}(t)$, we can find the Jost solution of
\eqref{eq:ZS} for $\tilde{q}_{ \scriptscriptstyle R}(t)$. 
Assuming $T>t_R=t_0$, 
$q_{ \scriptscriptstyle R}(t)\approx\tilde{q}_{ \scriptscriptstyle R}(t)$ if $t\geq T$ and zero, otherwise.
Finally, one can show for multi-soliton in Lemma~\ref{lem:symmetric_sol} that if $|\text{Im}(\lambda)|<\sigma_1$,
\begin{equation}\label{eq:right_scattering}
(a_{ \scriptscriptstyle R}(\lambda),b_{ \scriptscriptstyle R}(\lambda))
\approx
(\alpha(\lambda),e^{+2j\phi}\beta^*(\lambda^*))
\end{equation}
\begin{remark}
Another way to conclude \eqref{eq:right_scattering} is to find
the Jost coefficients of $q_{ \scriptscriptstyle R}(-t)$ in a way similar to the ones of $q_{ \scriptscriptstyle L}(t)$. Then,
$(a_{ \scriptscriptstyle R}(\lambda),b_{ \scriptscriptstyle R}(\lambda))$
are obtained
from the fact that the Jost coefficients 
of $q_{ \scriptscriptstyle R}(-t)$ are
$(a_{ \scriptscriptstyle R}^*(-\lambda^*),b_{ \scriptscriptstyle R}(-\lambda))$.
\end{remark}

\subsubsection{Layer-Peeling} Since $q(t)$ is decomposed into $q_{ \scriptscriptstyle L}(t)$, $q_{ \scriptscriptstyle T}(t)$ and $q_{ \scriptscriptstyle R}(t)$ with disjoint supports, 
the layer-peeling method~\cite{yousefi2014information} relates the Jost coefficients as follows:
\begin{align*}
a(\lambda)&=a_{ \scriptscriptstyle L}(\lambda)a_{ \scriptscriptstyle T}(\lambda)a_{ \scriptscriptstyle R}(\lambda)
-b_{ \scriptscriptstyle L}(\lambda)b^*_{ \scriptscriptstyle T}(\lambda^*)
a_{ \scriptscriptstyle R}(\lambda)\\
&\hspace{1cm}
-a_{ \scriptscriptstyle L}(\lambda)b_{ \scriptscriptstyle T}(\lambda)
b^*_{ \scriptscriptstyle R}(\lambda^*)
-b_{ \scriptscriptstyle L}(\lambda)a^*_{ \scriptscriptstyle T}(\lambda^*)b^*_{ \scriptscriptstyle R}(\lambda^*)\\
b(\lambda)&=a_{ \scriptscriptstyle L}(\lambda)a_{ \scriptscriptstyle T}(\lambda)b_{ \scriptscriptstyle R}(\lambda)
-b_{ \scriptscriptstyle L}(\lambda)b^*_{ \scriptscriptstyle T}(\lambda^*)
b_{ \scriptscriptstyle R}(\lambda)\\
&\hspace{1cm}
+a_{ \scriptscriptstyle L}(\lambda)b_{ \scriptscriptstyle T}(\lambda)
a^*_{ \scriptscriptstyle R}(\lambda^*)
+b_{ \scriptscriptstyle L}(\lambda)a^*_{ \scriptscriptstyle T}(\lambda^*)a^*_{ \scriptscriptstyle R}(\lambda^*)
\end{align*}
We replace \eqref{eq:left_scattering} and \eqref{eq:right_scattering} in the above equations.
Let define
\begin{align*}
\tilde{a}(\lambda)&=\alpha^*(\lambda^*)\beta^*(\lambda^*)a(\lambda)e^{j\phi},\hspace{0.3cm}\tilde{b}(\lambda)= b(\lambda)e^{-j\phi}\\
\tilde{a}_{ \scriptscriptstyle T}(\lambda)&=\alpha(\lambda)\beta^*(\lambda^*)a_{ \scriptscriptstyle T}(\lambda)e^{j\phi},\hspace{0.25cm}
\tilde{b}_{ \scriptscriptstyle T}(\lambda)= b_{ \scriptscriptstyle T}(\lambda)e^{-j\phi}
\end{align*}

Using the above relations in the equations of $a(\lambda)$ and $b(\lambda)$ and the fact that
$\alpha(\lambda)\alpha^*(\lambda^*)+\beta(\lambda)\beta^*(\lambda^*)=1$, we can derive the four following equations resulting \eqref{eq:a_lam} and \eqref{eq:b_lam}
\begin{align*}
\tilde{a}_{ \scriptscriptstyle T}(\lambda)-\tilde{a}^*_{ \scriptscriptstyle T}(\lambda^*)
&= \tilde{a}(\lambda) - \tilde{a}^*(\lambda^*)\\
\tilde{a}_{ \scriptscriptstyle T}(\lambda)+\tilde{a}^*_{ \scriptscriptstyle T}(\lambda^*)
&=\left(\alpha(\lambda)\alpha^*(\lambda^*)-\beta(\lambda)\beta^*(\lambda^*)\right)\left(\tilde{a}(\lambda)+\tilde{a}^*(\lambda^*)\right)\\
&\hspace{0.4cm}+2\alpha(\lambda)\alpha^*(\lambda^*)\beta(\lambda)\beta^*(\lambda^*)
\left( \tilde{b}(\lambda)+\tilde{b}^*(\lambda^*)  \right)\\
\tilde{b}_{ \scriptscriptstyle T}(\lambda)+\tilde{b}^*_{ \scriptscriptstyle T}(\lambda^*)
&=\left(\alpha(\lambda)\alpha^*(\lambda^*)-\beta(\lambda)\beta^*(\lambda^*)\right)\left(\tilde{b}(\lambda)+\tilde{b}^*(\lambda^*)\right)\\
&\hspace{0.4cm}-2\left(\tilde{a}(\lambda) + \tilde{a}^*(\lambda^*)\right)\\
\tilde{b}_{ \scriptscriptstyle T}(\lambda)-\tilde{b}^*_{ \scriptscriptstyle T}(\lambda^*)
&= \tilde{b}(\lambda) - \tilde{b}^*(\lambda^*)
\end{align*}

\section{Conclusion}\label{sec:conclusion}
To increase the spectral efficiency and data rate,
multi-soliton pulses are in practice truncated before transmission over the optical fiber.
The truncation distorts slightly the nonlinear Fourier spectrum of a multi-soliton pulse.

In this paper, we derived simple closed-form expressions for the nonlinear Fourier spectrum when the tails of a symmetric multi-soliton are truncated. Although the analysis has some tight approximation of the tails, the closed-form expressions, for both continuous spectrum and the discrete spectrum, seem to be very precise  when the truncation is not very aggressive. 

We further showed how to find the eigenvalues of the discrete spectrum from the continuous spectrum. We presented this general method for the application of this paper.

Although we presented the results for symmetric multi-solitons but the analysis is more general and can be extended to any multi-soliton pulses with similar tails behaviour.

\bibliographystyle{IEEEtran}
\small
\bibliography{references}

\end{document}